\documentclass[11pt,a4paper]{article}
\pdfoutput=1 
\usepackage{graphicx}
\usepackage{units}
\usepackage[top=1.2in, left=2cm, right=2cm, includefoot]{geometry}
\usepackage{fancyhdr}
\usepackage{amsmath}
\usepackage{graphicx}
\usepackage{bm}
\usepackage{svg}
\usepackage{appendix}
\usepackage{mathtools}
\usepackage{authblk}

\usepackage{mathtools}

\newcommand{\pderiv}[2]{\frac{\partial#1}{\partial#2}}
\newcommand{\pdd}[3]{\frac{\partial^2#1}{\partial#2\partial#3}}

\renewcommand{\d}[1]{\ensuremath{\operatorname{d}\!{#1}}}
\newcommand{\x}{\mathbf{x}}
\newcommand{\z}{\mathbf{z}}
\newcommand{\deriv}[2]{\frac{\text{d}#1}{\text{d}#2}}

\DeclareMathOperator{\Diag}{Diag}
\DeclareMathOperator{\Tr}{Tr}
\newcommand{\defeq}{\vcentcolon=}

\def\be{\begin{equation}}
\def\ee{\end{equation}}     
\def\bfi{\begin{figure}}
\def\efi{\end{figure}}
\def\bea{\begin{eqnarray}}
\def\eea{\end{eqnarray}}

\usepackage{mathtools}
\newlength{\arrow}
\newcommand*{\myrightarrow}[1]{\xrightarrow{\mathmakebox[\arrow]{#1}}}
\usepackage{amssymb}

\begin{document}

\title{Survival of the densest accounts for the expansion of mitochondrial mutations in ageing}

\author[1]{Ferdinando Insalata}
\author[1]{Hanne Hoitzing}
\author[1]{Juvid Aryaman}
\author[1,2, *]{Nick S. Jones}

\affil[1]{Department of Mathematics, Imperial College London, London, SW7 2AZ, United Kingdom}
\affil[2]{EPSRC Centre for the Mathematics of Precision Healthcare, Imperial College London, London, SW7 2AZ, United Kingdom}
\affil[*] {Corresponding author: nick.jones@imperial.ac.uk}
\date{}                  

\maketitle


\begin{abstract}
\noindent
The expansion of deleted mitochondrial DNA (mtDNA) molecules has been linked to ageing, particularly in skeletal muscle fibres; its mechanism has remained unclear for three decades. Previous accounts  assigned a replicative advantage to the deletions, but there is evidence that cells can, instead, selectively remove defective mtDNA. We present a spatial model
that, without a replicative advantage, but instead through a combination of enhanced density for mutants and noise, produces a wave of expanding mutations with wave speed consistent with experimental data, unlike a standard model based on replicative advantage. We provide a formula that predicts that the wave speed drops with copy number, in agreement with experimental data. Crucially, our model yields travelling waves of mutants even if mutants are preferentially eliminated. Justified by this exemplar of how noise, density and spatial structure affect muscle ageing, we introduce the mechanism of stochastic survival of the densest, an alternative to replicative advantage, that may underpin other phenomena, like the evolution of altruism. 
\end{abstract}

\section*{Introduction} 
\label{intro}
Mitochondria are central in health and disease  \cite{Wallace05}. 
Mutations in the mitochondrial DNA (mtDNA) can impair
organellar functions \cite{Tuppen10}, with negative impact on cellular physiology. 
These mutations can be inherited or  acquired over time \cite{Aryaman19_het}. The accumulation of DNA mutations to high levels has been repeatedly linked to ageing  \cite{Harman72, Krishnan07, Kauppila17}, 
especially in postmitotic tissues such as neurons or muscles \cite{Sun16}. In skeletal muscle, when the proportion of mutants in a zone of the fibre exceeds a threshold value \cite{Rossignol03, Aryaman17_biochem}, an oxidative phosphorylation defect is triggered, that is often revealed through deficiency in cytochrome \textit{c} oxidase activity. Skeletal muscle fibre damage occurring with age leads to a reduction in muscle mass and strength through 
atrophy and fibre breakage, termed sarcopenia \cite{Rolland08}. 
This is one of the first manifestations of old age  and has a knock-on effect on general health. Sarcopenia in mammals has been connected to high levels of mitochondrial deletions in skeletal muscle fibres \cite{Ferri20, Bua02, Herbst07, Alway17}. 

The high levels of mitochondrial deletions observed in ages skeletal muscle are attributed less to a high mutation rate than to the \textit{clonal expansion} of these mutations. Indeed, clonality is the defining feature of this expansion. Single-cell studies have consistently reported that fibres in damaged zones of the muscles are taken over  by a single deletion. This has been observed
in rats \cite{Cao01, Herbst07}, rhesus monkeys \cite{Gokey04, McKiernan09} and humans \cite{Brierley98}.
The fact that different affected fibres generally have different deletions suggests 
that founder mutations occur during the lifetime of an individual and then expand clonally \cite{Brierley98}. 
The mechanism behind this clonal expansion remains to be established,
and has eluded scientists for more than three decades 
\cite{Wallace89, Arnheim92, deGray97, Chinn99, Capps03, KDK14}.

In evolutionary theory, if one species (here the mtDNA deletions) 
invades a system and outcompetes another species (the wildtype mtDNA), it is often concluded that the former 
has a replicative advantage. This has been the approach followed by numerous authors \cite{KDK14, KK13, KK14, Lawless20, Tam15}, who have made extensive use of mathematical population genetics
modelling in order to compare quantitative predictions
with experimental data.
However,
there is no accepted mechanism to justify a replicative advantage for mtDNA deletions that can explain clonal expansion.
It was initially proposed that deletions, being smaller, replicate in a shorter time, leading
to a replicative advantage \cite{Wallace89, Wallace92}. 
This was observed in some specific cases, namely for 
fast-proliferating cells \cite{Russell18} and cells recovering their mtDNA populations after heavy
depletion \cite{Diaz02}. However, through numerical simulations it has been shown that the size-induced
advantage cannot explain the observed accumulation in non-replicating muscle fibres, 
especially in short-lived animals \cite{KDK14}. 

Other mechanisms have been proposed, termed the vicious cycle  \cite{Arnheim92}, 
survival of the slowest \cite{deGray97}, survival of the sick \cite{Lane11} 
and crippled mitochondria \cite{Yoneda92}, but 
they were later found implausible and/or incapable 
of accounting quantitatively for  experimental data (see Ref. \cite{KK18} for a review). Other studies  have modelled the phenomenon through a neutral stochastic model,
i.e. without assigning a replicative advantage to mutants \cite{Chinn99, Elson01, Capps03}.
In this approach, it is argued that random drift alone
is sufficient to explain the clonal expansion. However, a more recent study \cite{KK13} has shown that, for models like this, neutral random drift can account for clonal expansion only for very long-lived species (see penultimate section of Results).

In this work, we develop and test a mathematically transparent, physically-motivated stochastic model of the expansion of 
mtDNA deletions, that takes into account the spatial structure of muscle fibres as
multinucleate cells, with many nuclei distributed along their length, each one 
controlling  a surrounding population of mitochondria that 
can diffuse along the fibre (Fig.~\ref{first_fig}A). All previous models have treated muscle fibres as a single well-mixed region, 
containing an isolated and unstructured population of 
mitochondria. 

Our work establishes that mitochondrial deletions can expand in muscle fibres without a replicative
advantage and, surprisingly, even if mutants are preferentially eliminated from the system.
Despite being subject to higher rates of mitophagy -- something that finds experimental support 
\cite{Shadel97, Fernandez03, Twig08, Suen10, Kim11, Aryaman19_net} -- mutants can come to dominate zones of the fibres.
We also predict that the expansion of deletions takes place through  a \textit{travelling wave} of 
mutants propagating in the muscle fibres. We are able to recapitulate the spatial features of the
distribution  of the mutant load along fibres and align our quantitative predictions with available
experimental data, using a model with only 4 parameters (with each parameter being constrained by auxiliary experimental data).

Our model structure derives from the generalized Lotka-Volterra model \cite{Hofbauer98},
although our conclusions are robust to other model choices widely used in mitochondrial genetics.
We provide a mathematical expression for the mutant propagation wave speed, finding that it drops as 
the density of mtDNA in the fibre increases; we corroborate this prediction on 
further experimental data. In contrast to the model we present, we show that a standard model based on a replicative advantage of mutants 
produces an implausibly fast clonal expansion, incompatible with experimental data. Finally, a further quantity of interest is the rate of occurrence of founder mutations. We argue that previous neutral models, neglecting the spatial structure of the system, 
have greatly overestimated this rate. 

To the best of our knowledge, the
clonal expansion of mitochondrial deletions in skeletal muscle fibres 
is the first phenomenon with a compelling description
as stochastic \textit{survival of the densest}: a species that outcompetes another due to a combination of increased density, spatial structure and stochasticity, that can counterbalance a higher death rate.
We claim that this mechanism might be the driving force behind other counterintuitive 
evolutionary phenomena, such as the spread of altruistic behaviour, that has analogies to the expansion of deletions. 
Stochastic survival of the densest is separate from frequency and density-dependent selection (e.g. the Allee effect \cite{Allee27}),  and spatially-structured game theoretic models (see SI.6 for further discussion).

\section*{Results}

\subsection*{Mean heteroplasmy increases in a spatial stochastic model with preferential elimination of mutants}
In our full model, skeletal muscle fibres are schematised as a chain of regions, with each region
representing a system of a nucleus surrounded by a population of mtDNA molecules, subdivided into wildtype and mutants. Individual molecules of mtDNA can be degraded, replicate and diffuse along the muscle fibre. We begin by analysing a single region. We will then explore the consequences of allowing two regions to be coupled, allowing mtDNAs to hop between adjacent regions. Extended chains of such regions will be considered in the following section. Importantly, this paper focuses on the dynamics of a single pre-existing mutation that reaches high levels through clonal expansion. We thus \textit{do not consider de novo mutations} occurring continuously through time. 

\begin{figure}
\centering
\includegraphics[width=\columnwidth]{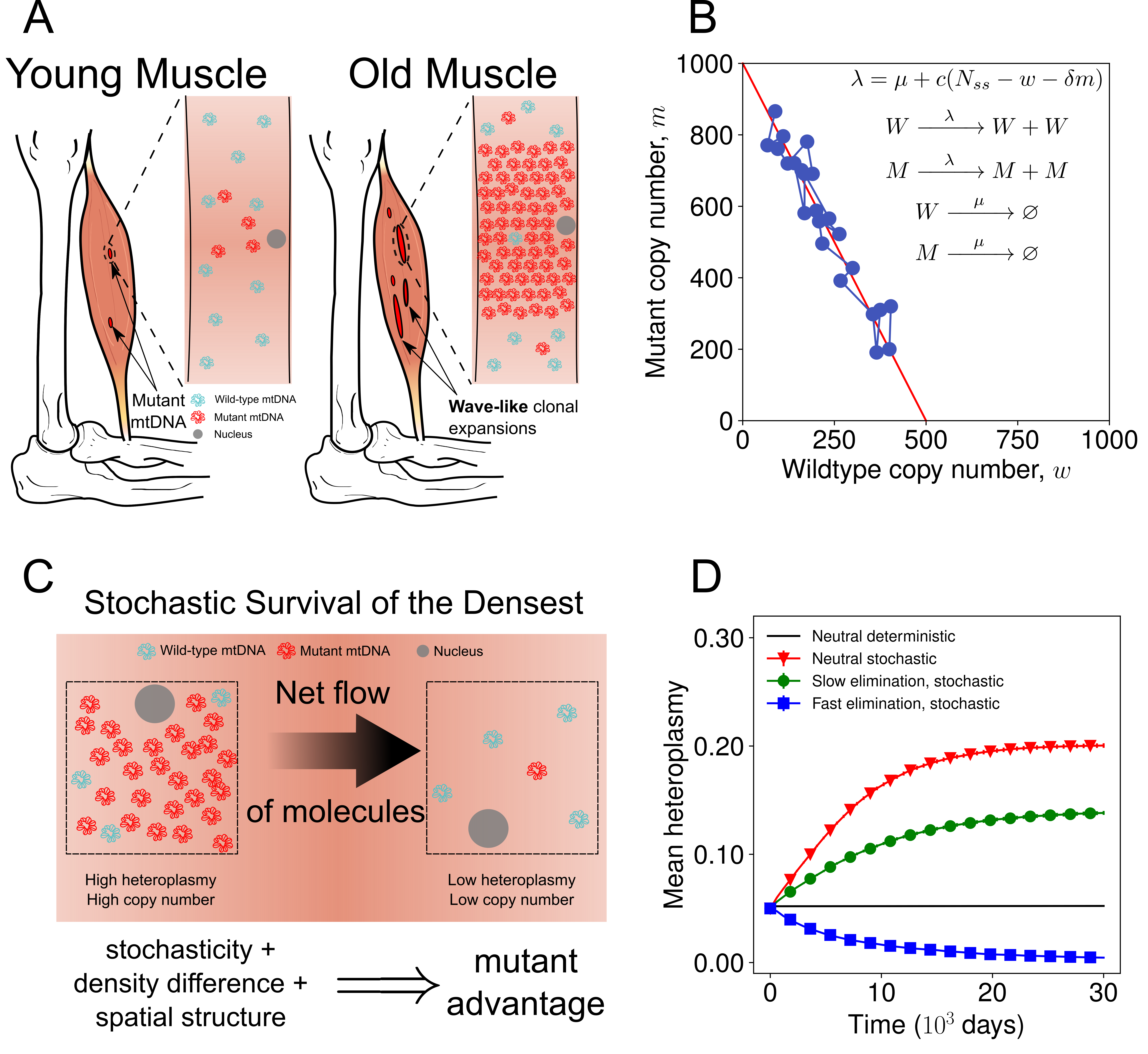}
\caption{
\textit{Stochastic survival of the densest can produce increases in the proportion of mutants even if they are subject to higher mitophagy rates than wildtypes.} 
(A) Dysfunctional mtDNA mutants expand in
muscle  fibres with age in a wave-like manner, leading to defects in OXPHOS. (B) The one-region system fluctuates around 
the steady state line (red line), on average moving toward regions of higher $m$ (average proportion of mutant $\langle h \rangle $ stays constant). 
(C) Differences in carrying capacity, noise and spatial structure (exchange of mtDNA between neighboring regions) lead to an increase in $\langle h \rangle$, an effect we have termed stochastic survival of the densest.
(D) In the stochastic 
two-region system, mean mutant fraction (heteroplasmy) increases in the neutral model (red)
and even if mutants are preferentially eliminated (green). Parameter values in SI.\ref{toy_params}. Error bars are SEM. }
\label{first_fig}
\end{figure}

For one region, we use the stochastic model
first introduced in Ref. \cite{Johnston16} and further analysed in Ref. \cite{Hoitzing19}, that is close to the established model \cite{Chinn99}.
We denote the wildtype and mutant copy number respectively by $w$ and $m$. Heteroplasmy is the mutant proportion, namely $h=m/(m+w)$.
The per-capita degradation rate $\mu$ is common and constant; the per-capita  replication rate $\lambda(w,m)$ is also common to both species, but state-dependent, and is
\begin{equation}
\lambda(w,m)= \mu + c(N_{ss}-w-\delta m),
\label{linear_f_c}
\end{equation}
with parameters $c, N_{ss}$ and $\delta$ interpreted below.
We remark that \textit{these birth and degradation rates are the same for mutants and wildtypes}: no explicit
selective advantage is assumed and the model can be called neutral.
The full model is formalised via a reaction network (Fig.~\ref{first_fig}B) and Eq.~\eqref{stoch_model}) as a Poisson point process.

The parameter  $\delta$ is one of the key elements of the model. Setting $\delta \neq 1$ allows for a differential contribution to the replication rate between mtDNA species \cite{Hoitzing19}. In SI.\ref{1_region_SI_det} we show that $\delta \neq 1$ is equivalent to having a difference in carrying capacity or \textit{density} between the two species and  that $0<\delta <1$, corresponds to mutants being the densest species. In the following it is always $0<\delta <1$.

After a transient, the system hits a state for which $\lambda = \mu$ and then starts fluctuating around the deterministic steady state line, of equation  $w + \delta m =N_{ss}$ in the $w-m$ plane, as depicted in Fig.~\ref{first_fig}B. Hence the parameter $N_{ss}$ can be interpreted as the target toward which the effective population $w + \delta m$ is steered.
Typical values of $N_{ss}$ in healthy human tissues are $10^3 - 10^4$
\cite{Miller03}. The parameter $c>0$ is a control strength (see SI.\ref{1_region_SI} for interpretation).

As a number of authors have observed \cite{Chinn99, Capps03, Elson01, Hoitzing19, Const17}, this model exhibits an increase in mean mutant copy number, in its neutral version defined in Eq.~\eqref{linear_f_c} and  even in the presence of preferential elimination of mutants in the form of a higher degradation rate. Intuitively, while fluctuating around the steady state line the system on average drifts toward regions of higher $m$ (Fig.~\ref{first_fig}B). We refer to SI.\ref{1_region_SI} for a mathematical treatment and an intuitive explanation of this effect that, as we show below and in the next section, lies at the core of our explanation of muscle ageing.

While an increase in mean mutant copy number is known, here we investigate another property: an \textit{increase in mean heteroplasmy} $\langle h \rangle $ in a spatially structured system, whose regions are coupled together and exchange mtDNA molecules.
We first couple together
two regions (Fig.~\ref{first_fig}C), with each region’s population evolving according to the dynamics described above and with the addition of exchange of mtDNA molecules at a constant rate per molecule $\gamma$,
an additional parameter whose value we derive from experimental studies (SI.\ref{parameters}).  
In this setting, in addition to the increase in mean mutant copy number
we also observe an \textit{increase in mean heteroplasmy}  $\langle h_i \rangle$ in each region, in a neutral model 
(Fig.~\ref{first_fig}D, red line), as well as in the presence of enhanced clearing of mutants (Fig.~\ref{first_fig}D, green line).
In the single region case, mean heteroplasmy is, by contrast, constant and always decreases when mutants are preferentially degraded (Fig.~\ref{fifth_fig}B).

Two key elements are needed for a spatially-extended system to exhibit the increase in mean heteroplasmy. First, noise: the effect is \textit{not} observed in the deterministic version of our model (Fig.~\ref{first_fig}D, black line), that we define in SI.\ref{2_regions_SI}. Second, mutants must be the densest species, i.e. $0<\delta <1$. Therefore, we have termed this effect \textit{stochastic survival of the densest} (SSD). A heuristic explanation is that the mean copy number of the densest species is driven up by stochastic fluctuations over time. Then, when we couple regions together allowing molecules of mtDNA to hop between adjacent regions, the local increase in mutant copy number causes a net flow of mutants into neighbouring regions (Fig.~\ref{first_fig}C). This, in turn, leads to an  increase in the mean \textit{proportion} of mutants with time.

\subsection*{Stochastic survival of the densest is favoured as a model for muscle fibre ageing}
\label{wavespeed}
In this section, we ask whether SSD, the effects highlighted in the previous section, can 
reproduce experimental data on the expansion of deletions in skeletal muscle fibres. We also compare
SSD to a standard model based on giving mutants a replicative advantage over wildtypes.
Coupling together hundreds of regions along a line we obtain a physical model 
of a skeletal muscle fibre. 

\begin{figure}
\includegraphics[width=\columnwidth]{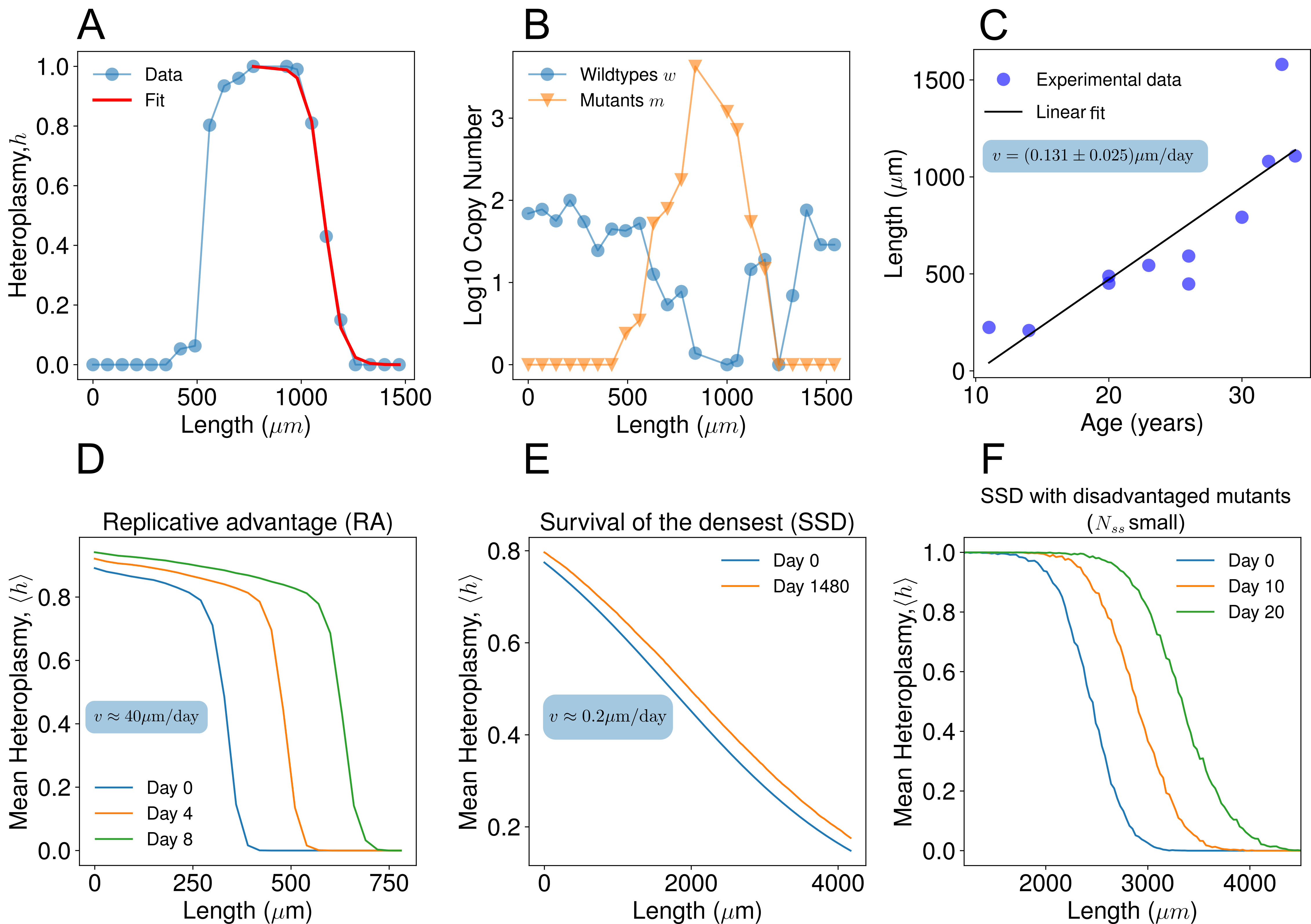} 
\centering
\caption{\textit{Stochastic survival of the densest predicts a wave-like expansion of mtDNA mutants at 
a speed in agreement with experimental observations, while a standard replicative advantage model predicts a speed a factor of $\approx 300$ too large.} 
A) Spatial profile of mutant fraction (heteroplasmy)
along a human skeletal muscle fibre \cite{Bua06}. The heteroplasmy profile follows a sigmoid, the shape expected for a travelling wave.
(B) Spatial structure of copy number for wildtype (blue) and mutants 
(orange) for the same muscle fibre as in panel A.
The heteroplasmic zones present a higher absolute copy number, i.e. mutants are present at a higher 
density. (C) Experimental data on the length of abnormal zones of muscle fibres against age in rhesus 
monkeys, from Ref. \cite{Lopez00}. An approximate linear relationship is found ($R^2=0.76, p=5 \cdot 10^{-4}$), compatible
with a wave-like expansion with speed $(0.131  \pm 0.025)\mu m/$day (linear fit). (D) Stochastic simulations of a spatially extended model with a replicative advantage for mutants, with our best estimate of the model parameters (see SI.\ref{parameters}) for muscle fibres of rhesus monkeys predicts a wave-like expansion with a speed of  $\simeq 40 \mu m/$day, 300 times
faster than the observed speed. (E) Simulations of  survival of the densest, with same death and replication rate for wildtypes and mutants, yields
a mutant wave speed of $\simeq 0.2 \mu m/$day for the fibres or rhesus monkeys, which is comparable with experimental observations (see (C)). (F) Survival of the densest predicts a wave of mutants even when mutants have a degradation rate higher than wildtypes and the same replication rate. Parameters are different from (E) and listed in SI.\ref{parameters}.}
\label{second_fig}
\end{figure}

Fig.~\ref{second_fig}A shows data obtained from human vastus lateralis samples \cite{Bua06}: 
a high-heteroplasmy
zone is present, flanked by transition zones to low or zero heteroplasmy. 
Previous mathematical models of mtDNA deletion expansion have not addressed this spatial structure.
A well-known feature of the clonal expansion of deletions is the high absolute copy number 
in zones of the muscle fibres taken over by mutants.
This is clearly visible in  Fig.~\ref{second_fig}B, which reports data on mtDNA copy 
number for the same muscle fibre \cite{Bua06}. In the zone where 
only mutants are present  the absolute
copy number is approximately 5 times higher than in the wildtype-only
zone, meaning a fivefold increase in carrying capacity for mutants. A fivefold increase 
is found also in Ref. \cite{Barthelemy01}, for a patient with ophthalmoplegia associated
with an mtDNA deletion
in muscle. 

The functional form of the 
heteroplasmy profile in Fig.~\ref{second_fig}A (red fitted line) is convincingly 
the form expected for a travelling wave (see SI.\ref{steepness}),
namely a sigmoid $1/(1+e^{\tau x})$ with $x$ being the position along the muscle fibre centred at the point where $h=1$ and $\tau=(2.464 \cdot 10^{-2} \pm 2 \cdot 10^{-5})\mu m$ (maximum likelihood estimation, MLE)
suggesting that \textit{the expansion is a wave-like phenomenon}.
We have estimated the speed of this wave analysing data from from Ref. \cite{Lopez00} on the length of
abnormal zones in rhesus monkeys and age of the subject.
Regressing the lengths against age  (Fig.~\ref{second_fig}C), 
we observe a relationship $(p=5 \cdot 10^{-4})$ which is approximately linear ($R^2=0.76$) and corresponds to an
average wave speed of $(0.131  \pm 0.025)\mu m/$day. In SI.\ref{sec:monkeys_speed} we further discuss 
the analysis of these data and explain why the true wave speed is likely 
to be marginally larger.

We consider two ways of modelling mutant expansion and reproducing the fivefold increase in mutant carrying capacity, one corresponding to a mutant mtDNA with a higher replication rate  -- replicative advantage (RA) model --
and the other to our SSD model.
A standard RA model reproduces the fivefold increase in carrying capacity
through a larger replication rate for mutants. 
In this model, the wildtype replication rate is
Eq.~(\ref{linear_f_c}) with $\delta=1$, and the mutant replication rate is the same but with $N_{ss}$ replaced by $5N_{ss}$ (see Eq.~\eqref{eq:mutant_repl_rate_RA}). 
The death rate  $\mu$  will be constant and common to the two species.
It has been known since the work of Fisher and Kolmogorov \cite{Fisher37, Kolmogorov37} 
that an advantageous mutation arising in a spatially extended system can come to dominate it,
sweeping it in a wave-like fashion, with a high-heteroplasmic front advancing at constant speed (FK waves henceforth). 
We have performed stochastic simulations of this  RA model, 
with values of the other parameters estimated from 
experimental data (SI.\ref{parameters}) and initialising the mutant content of the system as seen in
Fig.~\ref{second_fig}A. The simulations 
show a wave-like expansion with a speed of $40\mu m /$day (Fig.~\ref{second_fig}D), 300 times faster than 
the observed $0.131\mu m /$day. We interpret this as strong evidence against the hypothesis of a 
simple replicative advantage being the cause of the expansion of deletions in skeletal muscle fibres.  \textit{Therefore, a
standard replicative advantage model is not consistent with the observed spatiotemporal 
dynamics of the expansion of mtDNA deletions in muscle fibres}.

In our approach, SSD, 
we reproduce the increase in mutant carrying capacity, extending
our neutral model from the previous section  to a chain of regions, using  $\delta = 1/5$.
In Fig.~\ref{second_fig}E we show the results of the simulations for
our model, with all parameter values specified by the literature (SI.\ref{parameters}).
The predicted speed is $\simeq 0.2 \mu m/$day, 
an improvement of two orders of magnitude over the model based on replicative advantage.
It is exclusively the stochastic version
of the model that exhibits this wave-like expansion; in the deterministic version, the high-heteroplasmy 
front only diffuses away (Fig.~\ref{seventh_fig}A). 
Moreover,  \textit{stochastic survival 
of the densest reproduces the expansion of mutants even if they are preferentially degraded} (see Fig.~\ref{second_fig}F),
whereas in a deterministic model a higher degradation rate for mutants leads to their extinction (Fig.~\ref{seventh_fig}B).
The key ingredients for a spatially-extended system to show SSD
are thus stochasticity and differences in carrying capacity, or density, between the two species.

The RA model (FK waves)
predicts that wave speed increases when $N_{ss}$ increases \cite{Hallat11}. 
SSD predicts, on the contrary,
that a smaller $N_{ss}$ yields a faster wave. Intuitively, this is the case because 
the expansion of mutants is driven by stochastic fluctuations, 
whose  effect generally becomes smaller in larger systems (see also Eq. (\ref{mut_SDE})).
Simulations reported in SI.\ref{pheno_wave_speed} for the neutral model reveal how wave speed varies 
as a function of the parameters of our model and we find they can be approximated by the phenomenological equation
\begin{equation}
v \simeq 2  \sqrt{kD},
\label{wavesp_formula}
\end{equation}
with $k=k_{SSD}=\nicefrac{\sqrt{(1-\delta)^2 \mu \gamma}}{N_{ss}^\frac{2}{3}}$. 
Eq.~\eqref{wavesp_formula} is also the wave speed of a FK wave but where $k$ is the net replicative advantage $k_{RA}$. In our case $k_{SSD}$ is interpretable as the effective advantage induced by SSD.

\subsection*{Stochastic survival of the densest is consistent with wider data and the effect is robust to model and parameter variation}
\begin{figure}
\centering
\includegraphics[width=0.8\textwidth]{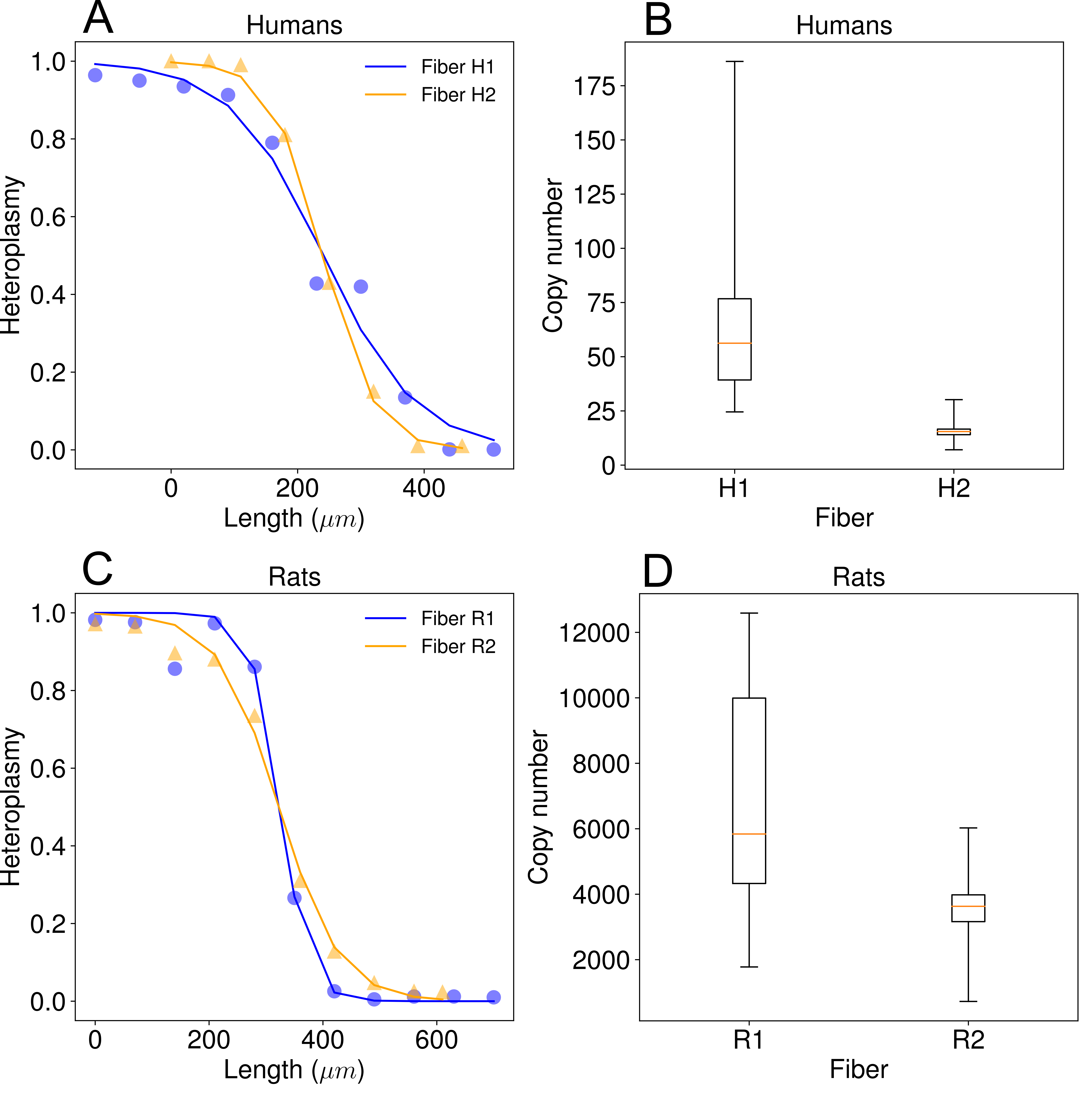}
 \caption{\textit{A steeper wave of mutants propagates more slowly in fibres which have a higher copy number,
in agreement with the predictions of stochastic survival of the densest}. 
(A) Significant difference in the steepness of wavefronts of two human muscle fibres H1 and H2: 
$\tau_{H1}=(2.4 \cdot 10^{-2} \pm 10^{-3})/\mu m$ for H1 and $\tau_{H2}=(1.4 \cdot 10^{-2} \pm 2 \cdot 10^{-3})/\mu m$ for H2  (MLE).
Data from Ref. \cite{Bua06}. According
to the mathematics of travelling waves (SI.\ref{steepness}), 
the steeper wave is slower. 
(B) Comparison between the corresponding $N_{ss}$ of the two fibres H1 and H2. 
We have found evidence ($p= 10^{-4}$, one-tailed Welch's \textit{t}-test) that the average 
copy number in normal zones of the two fibres H1 and H2, 
$N_{ss}$ in our model, is higher 
for the steeper, and hence slower, wave. 
This is in qualitative agreement with the predictions 
of our stochastic survival of 
the densest model that wave speed decreases with copy number. In contrast, a model based on replicative advantage predicts that speed increases with copy number. 
(C) Two muscle fibres in rats, R1 and R2,
present waves of mutants with significantly different steepness of
the waveform: $\tau_{R1}=(4.0 \cdot 10^{-2} \pm 6 \cdot 10^{-3} )/\mu m$ for R1 and $\tau_{R2}=(1.8 \cdot 10^{-2} \pm 2 \cdot 10^{-3}  )/\mu m$ for R2 (MLE). 
(D) We have found an indication ($p=0.06$, one-tailed Welch's \textit{t}-test) 
that the average 
copy number in normal zones of the two fibres R1 and R2, 
is higher for the steeper wave (R1).
Data from Ref. \cite{Herbst07}.}
\label{third_fig}
\end{figure}

The leading edge of a faster 
wave is flatter than that of a slower wave (e.g. Ref. \cite{Murray02}, summarised
in SI.\ref{steepness}).  
By exploiting this property, it is possible to compare
the speeds of two waves by examining their shapes.
Data on muscle fibres
for humans \cite{Bua06} (Fig.~\ref{third_fig}A, B) and rats \cite{Herbst07} (Fig.~\ref{third_fig}C, D) show that
flatter waves of mutants propagate along fibres with a smaller $N_{ss}$.
This is in agreement with the prediction of our model and against the predictions for FK waves.

Skeletal muscle fibres can be broadly classified into Type 1 (oxidative) and Type 2 (glycolytic) fibres
\cite{Ranvier73}.
The former rely on OXPHOS to function and have twice as many mitochondria as the the latter
\cite{He02,Ogata85, Lopez00}, that
depend on glycolysis.
It is known that Type 2 fibres are more affected by sarcopenia 
with ageing \cite{Bua02, Herbst16, Wanagat01, Larrson83, Picard11, Lopez00}. 
While there are other physiological differences between the fibre types, 
this is in line with the predictions of our model that a smaller 
copy number per nucleus produces a faster expansion of mutants. 
Small, short-lived animals like rodents 
show sarcopenia on a time-scale of years (from $\simeq$ 2 years)
and have a larger proportion of Type 2 fibres compared to long-lived animals such as
rhesus monkeys and humans
\cite{Pellegrino03}, that exhibit sarcopenia on a longer time-scale (decades).

Studies involving changes to mitochondrial copy number are consistent with an inverse relationship between copy number per nucleus and wave speed.
It has been found that copy number depletion caused by antiretroviral therapy \cite{Abdul18}
or AKT2 deficiency \cite{Chen19} is associated with enhanced sarcopenia. Likewise, statins are well known
for increasing the risk of sarcopenia \cite{Distasi10, Ramachandran17, Dabrowa18} and have consistently
been associated with reduction in mitochondrial copy number \cite{Paiva05, Stringer13, Cirigliano20}. 
Conversely, increase in mtDNA content through exercise \cite{Yoo18, Theilen17} 
or overexpression of  TFAM \cite{Theilen19}
and parkin \cite{LeducGaudet19} have been found to protect against sarcopenia and muscle atrophy.
If mtDNA deletions had a replicative advantage, 
the inverse relationship between copy number per nucleus and wave speed would not be observed.
 
\begin{figure}
\centering
\includegraphics[width=\columnwidth]{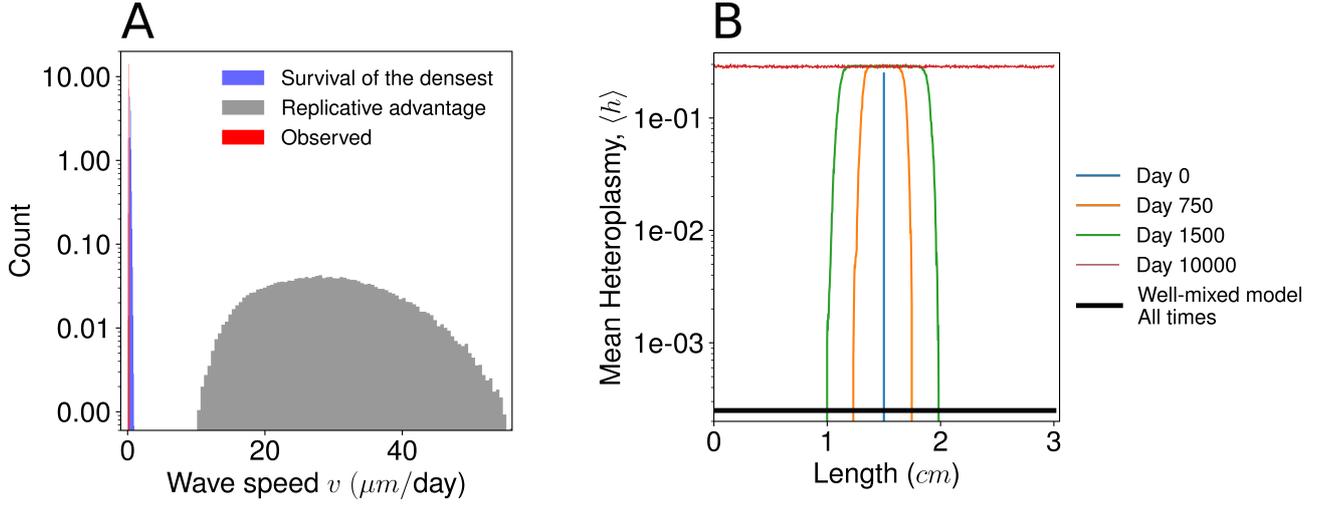}
\caption{\textit{Survival of the densest robustly predicts a wave speed two orders of magnitude closer to observations compared to a standard model based on replicative advantage, and predicts a greatly enhanced probability for the expansion of founder mutations than models lacking spatial structure.}
(A) Inserting probabilistic estimates of the model parameters 
(see SI.\ref{parameters}) into Eq.~\eqref{wavesp_formula} we find that 
survival of the densest predicts a distribution for the wave speed (red histogram) 
compatible with observations (blue), whereas a F wave driven by replicative advantage 
with the same model parameters is two orders of magnitude faster (grey).
(B) The probability that a single founder mutation (time 0, blue spike) takes over the system is equal to the mean heteroplasmy at long times in an ensemble of simulations. This probability dramatically increases in a spatially-structured system with $N$ regions ($ \approx 0.3$, red line) and $N_{ss}$ mtDNA molecules per region with respect to a single well-mixed region with the same total copy number $N N_{ss}$, which is how previous studies modelled skeletal muscle fibres \cite{Elson01, KK13, KK14}, where it stays constant at the initial value ($1/(N_{ss}N) \approx 2.5 \cdot 10^{-4}$, black line). Lines relative to previous times show the advance of the wave of heteroplasmy seeded by the founder mutation. Parameters are  $N=1001, N_{ss}=4, \delta=0.1, \mu=0.07/$day, $\gamma=0.1/$day, $ c=0.04/$day.}
\label{fourth_fig}
\end{figure}

We have obtained probability distributions for the wave speeds predicted by survival of the densest and a RA model (FK waves), by inserting draws from the distributions of parameter values (given in SI.\ref{parameters})
into Eq. (\ref{wavesp_formula}), with the appropriate interpretation of $k$ for the two models. The predicted distributions are plotted in Fig.~\ref{fourth_fig}A together with the distribution of the experimentally observed wave speed obtained via linear fit (Fig.~\ref{second_fig}C). After accommodating this parametric uncertainty, survival of the densest remains much superior to RA at reproducing the observed speed.
Finally, we have verified numerically that our results  are robust to a selection of model choices for 
the control mechanism on the mitochondrial population. 
The main feature of our model, the travelling waves of mutants whose 
speed decreases for larger  $N_{ss}$, does not require the specific linear dependence of replication rate 
on copy number of Eq. (\ref{linear_f_c}). Other neutral feedback controls are possible, as long as they
encode a larger mutant density. In SI.\ref{other_feedbacks} we show that 
a \textit{quadratic} (Fig.~\ref{seventh_fig}C)  and \textit{reciprocal} (Fig.~\ref{seventh_fig}D) feedback control  produce a wave of mutants qualitatively similar to that produced by linear feedback.

\subsection*{Wavelike spreading implies that low mtDNA deletion rates can nonetheless yield high mutant load}
\label{Mut_rates}

It has often been assumed that neutral stochastic models of mtDNA dynamics cannot explain the mutational load in skeletal muscle fibres of short-lived animals, due to a perceived incompatibility between the \textit{de novo} mutation rate of mtDNA polymerase and observed mutational loads \cite{KK13, Cao01,Herbst07, KK18}. This is one of the reasons that has motivated theorists to develop RA models.
However, previous attempts to understand mutant proliferation in muscle fibres have modelled the system as a bulk of well-mixed mtDNAs \cite{Elson01, KK13, KK14}. In Fig.~\ref{fourth_fig}B we show the evolution of a chain of regions with $Nss=1001$ initialized with a single mutant molecule of mtDNA. We observe a high-heteroplasmy front, of mean height $\simeq 0.30$, meaning that the mutants reach 100\% heteroplasmy  around $30\%$ of the time. The fixation probability of a founder mutation increases by three orders of magnitude compared to the well-mixed case (black line). As we argue more in detail in SI.\ref{ssod_short_lived}, \textit{this result dramatically lowers the  de novo mutation rate needed to account for the observed mutant loads}, since a higher fixation probability implies that fewer mutations need to arise. Additionally, in SI.\ref{revisited_loads}, we argue that previous modelling studies overestimated mutant loads -- and therefore mtDNA mutation rates -- by using bulk models to interpret experimental data.
In conclusion, we have found two reasons why the mutation rate required to reproduce the observed mutant loads in aged skeletal muscle fibres without an RA for mutants is much lower than previously thought.
\subsection*{Implications for the evolution of altruism}
\label{darwin_wrong}
An altruistic trait is one that benefits others while costing its carriers \cite{Kreft04}.
In SI.\ref{1_region_SI_K}, specifically in Eq.~(\ref{eq:K_and_h}), we show that the global carrying capacity 
of the system -- the population it sustains -- increases with the proportion of mutants. In Eq.~\eqref{eq:lambda_and_h} we show that, equivalently,
the replication rate of both wildtypes and mutants is 
an increasing function of the proportion of mutants. 
If we assign a higher degradation rate to mutants, as we do in order to model a notional higher mitophagy rate
for mitochondrial deletions, then mutants can be considered an altruistic species. Indeed, the increase in carrying capacity and replication rate can be seen as a benefit that mutants bring to the whole population, including wildtypes, at the cost of a higher degradation rate.
Our model of the spread of mitochondrial deletions in skeletal muscle fibres can therefore be reinterpreted as a model for the spread of altruistic behaviour.  
 An increase in carrying capacity 
is present also in mathematical models of public good production \cite{Brockhurst08, Const16}  and cooperative use of limited resources \cite{Kreft04}. 
These two behaviours are
observed in microbial communities and are believed
to be early examples of altruism in the history of life \cite{Pfeiffer01}.


The two most prominent accounts for the evolution of altruism are kin
selection and group selection \cite{Uyenoyama80}. 
The former approach argues that altruism directed towards genetically
related individuals can increase the overall success of the altruistic gene, 
which is likely to be shared by relatives of the carriers \cite{Griffin02}. 
In group selection, altruism spreads 
through advantages conferred to the group; for instance, a group whose members are ready to sacrifice 
themselves for other members will likely prevail over other groups \cite{Darwin71}.
These approaches seek a 
\textit{deterministic} advantage for altruistic traits.
In our model the spread of altruism is \textit{driven by noise} and has no deterministic counterpart. We now highlight our contributions in this area, referring to SI.\ref{our_differences} for further links to the wider evolutionary biology literature.

A first, important merit of our model consists in describing the spread of an altruistic trait in
a spatially extended system using a widely-used model of population genetics,
the Lotka-Volterra model, which uses linear per molecule transition rates. The Moran model is another fundamental model of stochastic population genetics, 
dealing with the less general case of finite fixed-size populations.
Previously, the  study in Ref. \cite{Houch12} 
has shown that in a modified Moran Model with nonlinear rates, 
where the assumption of 
fixed population size is relaxed, altruists with a higher carrying capacity 
can have a higher fixation probability despite a higher death rate.
Analogous results are presented in Ref. \cite{Houch14}, which is based on a 
modification of the Wright-Fisher model,  
retaining the feature of discrete, non-overlapping populations.
Our continuous-time model uses simpler (linear) rates, and the 
relevance of the Lotka-Volterra model, widely used in population genetics, indicates 
that SSD might be a widespread effect.

The increase in mean mutant copy number in the single-region system 
in a neutral model was highlighted 20 years ago in the field of 
mitochondrial biology, in the context of the relaxed replication model
formulated in Refs. \cite{Chinn99, Elson01, Capps03, Hoitzing19} and further analysed in Refs. \cite{Johnston16, Hoitzing19}.
Models yielding an increase in mean \textit{copy number} of an altruistic species have found that space can amplify the effect of randomness \cite{Const16, Const17, Parsons17}, in the sense
that the altruistic species can tolerate a higher selective elimination in a spatially structured system.
A key contribution of our work is to show explicitly that spatial structure can lead 
to a qualitatively different result, namely an increase in the \textit{proportion} of altruists, that in turn leads to a travelling wave of altruists (the mutants in the mitochondrial setting). 

Another contribution of our study is the use of a physically motivated and biologically interpretable
microscopic  mechanism.
The study in Ref. \cite{Hallat11} presents a phenomenological, macroscopic model that exhibits a noise-driven wave of altruists. We show in the SI.\ref{chain_SI} that the model in Ref. \cite{Hallat11} is a 
special case of ours, in the limit of continuous space. 
We have further been able to obtain a formula for the wave speed covering 
a range of biologically and clinically relevant regimes not covered by other studies, namely 
without taking $\delta \rightarrow 1$ and without imposing the quasi-static limit, i.e. very slow diffusion (see SI.\ref{pheno_wave_speed}).

Finally, but perhaps most importantly, we have identified the mtDNA populations in 
skeletal muscles as a candidate system that shows a travelling wave of a preferentially eliminated species. 
Muscle ageing could therefore be the first example of a new class of evolutionary phenomena best described as SSD.

\section*{Discussion}
In this study, we have modelled the accumulation of mutation during ageing through a bottom-up, physically interpretable stochastic model based on Lotka-Volterra dynamics, one of the simplest classic models of population genetics. Our aim  has been to  explain why mitochondrial deletions  clonally expand despite possibly being preferentially eliminated through quality control mechanisms \cite{Shadel97, Fernandez03, Twig08, Suen10, Kim11}. Ours is the first spatial model to account for the spread of mutant mtDNA, schematising skeletal muscle fibres as a chain of nuclei each surrounded by a population of mitochondria that can move diffusively along the fibre.

The main achievement of our study is predicting a \textit{wave-like clonal expansion of deletions  even if they are subject to preferential elimination}. This  counterintuitive result has its roots in the increased density of deletions -- a widely observed fact -- and in the stochastic nature of the model.  Previous studies have tried to account for the expansion of deletions assigning them higher replication rate: we have discovered that a standard model with spatial structure and with a replicative advantage for mutants produces a wave of deletions that is 300 times faster than observed. Crucially, our model predicts, instead, a speed that is of the same order of magnitude as the observed one. We have accounted for this effect mathematically and shown that our observations are robust to the particular choice of replication rate for both species and to the uncertainty in parameter values. We have provided a candidate functional form for the wave speed  (Eq.~\eqref{wavesp_formula}) that has implications for therapy,  since existing drugs allow us to modulate some of the parameters influencing the propagation of mutants (i.e. copy number $N_{ss}$ and turnover rate $\mu$). Moreover, we have corroborated our prediction of a wave speed decreasing with copy number by examining how copy number changes  onset time and likelihood of developing sarcopenia. Eq.~\eqref{eq:general_effective_eps} highlights that the largest preferential elimination $\epsilon_M$ that mutants can tolerate and still expand also decreases with $N_{ss}$. Therefore a larger copy number not only slows down the clonal expansion, but also makes it more likely to be stopped by a given rate of selective mitophagy.
Finally, we have shown that the wave-like spread, together with a re-evaluation of the current estimates  of the mutant loads, greatly lowers the \textit{de novo} mutation rate needed to reproduce the observed clonal expansion. A travelling wave of mutant implies that a lower mutation rate can nonetheless yield pathology. Future work might explore the interaction between the model presented here and mitochondrial network dynamics, as done in Ref.~\cite{Tam15} for a model based on replicative advantage. Given the evidence for sharing of mtDNA between distinct cells \cite{Spees06, Berridge16, Jackson16}, the effect we identify need not be restricted to multinucleate muscle fibres but could  yield apparently invasive waves of mtDNA deletions in other tissue types.

In conclusion, our core claim is that it is not a replicative advantage, but the increased density of deletions and the randomness intrinsic to biological processes that drive the clonal expansion of mitochondrial deletions in skeletal muscle fibres. To our knowledge, this phenomenon is the first experimental candidate for what we have termed stochastic survival of the densest. The mechanism we propose might have applications elsewhere. For example, a classic setting for the wave-like spread of a trait is in the uptake of agriculture \cite {Ammerman71, Ammerman73, Ammerman79, Pinhasi05}. However agriculture might not impart an explicit replicative advantage, and might lead to higher death rates \cite{Robson10, Lambert09, Cohen89, Armelagos84}, but nonetheless spread due to an increase in carrying capacity of the land. We believe that this study and the simplicity of the microscopic model we use might pave the way for an increased recognition of this intriguing mechanism in evolutionary biology, in light of the connections we have drawn with the evolution of altruistic behaviour.


\section*{Acknowledgement}
The authors acknowledge Judd Aiken and Allen Herbst for advising on skeletal muscle
fibres data, Sam Greenbury and Iain Johnston for useful comments about the manuscript.
NJ acknowledges the Leverhulme Trust RPG-2019-408 and EPSRC EP/N014529/1.
FI is supported by Imperial College's President's PhD Scholarship.


\newpage
\section*{Supporting Information}\label{sec:Methods}

\setcounter{equation}{0}
\renewcommand{\theequation}{S\arabic{equation}}
\renewcommand\thefigure{S\arabic{figure}}
\setcounter{figure}{0}
\renewcommand\thetable{S\arabic{table}}
\setcounter{table}{0}
\setcounter{enumiv}{0}

\section{One-region model: deterministic and stochastic treatment}
\label{1_region_SI}

The fundamental unit of our model is the single region hosting a well-mixed
population evolving according to linear feedback control. This constitutes the building 
block of our description of the skeletal muscle fibre. 
Our physical model of  muscle fibres is a chain of these regions.

In our model, there are two species: wildtypes $w$ and and mutants 
(deletions) mitochondrial DNA molecules $m$.

Every molecule is degraded at rate $\mu$ and replicates at rate $\lambda(w,m)$ given by
\begin{equation}
\lambda(w,m)=max[0, \mu + c(N_{ss}-w-\delta m)],
\label{linear_f_c_SI}
\end{equation}
i.e. the larger between 0 and $ \mu + c(N_{ss}-w-\delta m)$. However, for biologically relevant values of the parameters, $ \mu + c(N_{ss}-w-\delta m)$ is practically always positive, as we noticed in our simulations. Therefore we can also use the simpler form in Eq.~\eqref{linear_f_c}. Because of the linear relationship between the replication rate $\lambda$ and the copy number we name the replication rate \textit{linear feedback control} .
 
The meaning of the parameters $N_{ss}$ and $\delta$ are explained in the main text, where we also mentioned that 
$c>0$ modulates the strength of the control. This means that the larger $c$, 
the more strongly the system is penalised to be away from steady state when $w+\delta m \neq N_{ss}$,
i.e. when the terms in parentheses on the right-hand-side (RHS) is  $\neq 0$ and hence $\lambda \neq \mu$.
In other words, a larger value of $c>0$ will cause a stronger push toward steady state.

In the following we study a deterministic and stochastic version of the model defined by the above rates.

\subsection{Deterministic (ODE) treatment}
\label{1_region_SI_det}

In a deterministic setting, the above rates give rise to the following system of ODEs:

\begin{equation} 
\begin{split}
\deriv{w}{t}&= cw(N_{ss}-w-\delta m) \\
\deriv{m}{t}&= cm(N_{ss}-w-\delta m). 
\end{split}
\label{ODE_system}
\end{equation}
The analysis of this deterministic dynamical system is useful to understand the  stochastic version of the model. We limit our analysis to $w>0, m>0$ 
given the meaning of the variables as copy numbers.
We start by noticing the existence of the trivial, unstable fixed point $(0,0)$.
We then notice a set of attractive fixed points which form the straight line of
equation $w~+~\delta m ~=~N_{ss}$ in the $(m,w)$ plane. 
We refer to this as the central manifold (CM) or steady state ($\mu= \lambda$) line.
This line includes the configurations in which there is only a species, namely
the wildtype fixed point $(N_{ss},0)$ and the mutant fixed point $(0,N_{ss}/\delta)$. 
The \textit{carrying capacity} of a species is the population size of this species at its fixed point, at which the species exists in isolation. We see that in our model the carrying capacities are  $N_{ss}$ for wildtypes and $N_{ss}/\delta$ for mutants. 
When  $0<\delta <1$ -- that is always the case in this work -- the mutant carrying capacity is  larger than wildtype. Another interpretation of the condition $0<\delta <1$ is that mutants are less tightly controlled or sensed by the system, as 
a change in the number of mutants entails a smaller variation in  $\lambda(w,m)$
than an equal change in the number of wildtypes. Hence, in our model mutants are the \textit{densest} and \textit{less tightly controlled} species.
The red line in Fig.~\ref{first_fig}B is the CM for a system
with $N_{ss}=500$ and $\delta = 0.5$. 

We define the \textit{heteroplasmy} $h$ as the proportion of mutants
\begin{equation}
h=\frac{m}{m+w},
\label{heteroplasmy}
\end{equation}
a conserved quantity of the dynamical system, namely 
\begin{equation}
\frac{d h}{dt}=0,
\label{conserved_h}
\end{equation}
which can be verified by direct calculation.
Eq. (\ref{conserved_h}) is equivalent to the ratio $m/w$ being constant,
meaning that any line passing through the origin is a constant-heteroplasmy line.
Hence, a way to summarise the behaviour of the system is: 
\begin{itemize}
\item If the system is in steady state, i.e. if $w~+~\delta m ~=~N_{ss}$, the dynamics stop.

\item If the system is not in steady state, it will move toward the CM along a line connecting the 
initial condition to the origin. 
\end{itemize}
 
The same behaviour can be recovered from the full solution of the system, which is

\begin{equation} 
\begin{split}
w(t)&=N_{ss} \frac{w_0}{(N_{ss}-w_0-\delta m_o)e^{-cN_{ss}t}+ w_0 + \delta m_0}  \\
m(t)&=N_{ss} \frac{m_0}{(N_{ss}-w_0-\delta m_o)e^{-cN_{ss}t}+ w_0 + \delta m_0}, 
\end{split}
\label{ODE_sol}
\end{equation}
where $w_0=w(0)$ and $m_0=m(0)$ are the initial conditions. From Eq. (\ref{ODE_sol}) it follows that
\begin{equation} 
w(t)+\delta m(t)=N_{ss} \frac{w_0 +\delta m_0}{(N_{ss}-w_0-\delta m_o)e^{-cN_{ss}t}+ w_0 + \delta m_0}.
\end{equation}
For $t \rightarrow \infty$, as 
$e^{-cN_{ss}t}\rightarrow 0$, $w(t)+\delta m(t) \rightarrow N_{ss}.$ This shows that the parameter $c$ is 
connected to how fast the state of the system decays to the CM, justifying its interpretation of 
$c$ as control strength.

In summary, in the deterministic single-region model the system evolves toward the
CM of equation $w~+~\delta m ~=~N_{ss}$  with the condition $\dot{h}=0$. 
We remark on the difference in carrying capacity of the two species, being modulated  by the parameter $\delta$.

\subsection{Carrying capacity of the whole system and replication rate increase with heteroplasmy}
\label{1_region_SI_K}
The global carrying capacity $K$ of the system is the value of the total population $n=w+m$ at steady state.
$K$ can be expressed as a function of $h$. Since $m=hn$ and $w=(1-h)n$ the steady state condition
$w~+~\delta m ~=~N_{ss}$ becomes
\begin{equation} 
(1-h)n+\delta hn = (1-h)K+\delta hK = N_{ss},
\end{equation}
where the first equality holds because we have defined $K$ as the value of $n$ at steady state.
This leads to
\begin{equation}
K \equiv K(h) = \frac{N_{ss}}{1-(1-\delta)h}.
\label{eq:K_and_h}
\end{equation}
For $\delta <1$, $K(h)$ is an increasing function of $h$: the higher the proportion of mutants, the larger 
the population that the system can sustain. When $h=1$, i.e. mutants are in isolation, the mutant carrying capacity $\frac{N_{ss}}{\delta}$ is recovered.
One way to interpret this is that mutants are using the resources of the system in a 
more economical way. The economical use of a limited resource is considered one of the earliest and simplest forms of altruism \cite{Kreft04, Hallat11}, that brings benefits to all the other individuals in the system regardless of their identity.

Another way of seeing the benefit that mutants bring to all the other molecules in the system is in terms of enhanced replication rate. By expressing $w,m$ in terms of $h,n$ as above, Eq.~\eqref{linear_f_c_SI} can be written  as 
\begin{equation}
\lambda (h,n) = \mu + c[N_{ss} - (1+h - \delta h)n],
\label{eq:lambda_and_h}
\end{equation}
that shows that, for a given population size $n$, the common replication rate is 
an increasing function of $h$ when $\delta <1$. Notice that here, differently from Eq.~\eqref{eq:K_and_h}, $n \neq K$ since we are not assuming that the system is in the steady state.

In the study of the expansion of mitochondrial deletions, mutants are assigned a higher degradation rate, to model the higher mitophagy rate to which they are subject.  The mitochondrial deletions of our model, for which $\delta <1$, can hence be seen in a very general sense as agents that benefit others paying a cost, in the form of a higher degradation rate. This aligns with the definition of biological altruism \cite{Houch12, Kreft04}, hence the mutants of our model can be considered an altruistic species.

\subsection{A stochastic treatment exhibits selection reversal}
In the stochastic formulation of the model, the \textit{per capita} 
degradation and replication rates $\mu$ and $\lambda$ are interpreted 
as instantaneous probability (in an infinitesimal time interval) that each  molecule is degraded or replicates. In this setting,
$w,m$ and $h$ are random variables, and we ask questions about their probability distributions and their 
moments.
Stochastic population dynamics models can be formalised as chemical reactions networks,
consisting, in the terminology of chemical systems, of a set of reactants, reactions and products. 
Consider a general chemical system consisting of $N$ distinct chemical species ($X_i$) interacting via $R$ chemical reactions, where the $j$\textsuperscript{th} reaction is of the form
\begin{equation}
s_{1j}X_1+\dots+s_{Nj}X_N \xrightarrow{\hat{k}_j} r_{1j} X_1+\dots+r_{Nj}X_N \label{eq:gen_chem_sys}
\end{equation}
where $s_{ij}$ and $r_{ij}$ are stoichiometric coefficients. We define $\hat{k}_j$ as the per molecule rate for the $jth$ reaction. A chemical reaction network can be described by the associated $ N \times R$
stoichiometry matrix  $S_{ij} = r_{ij}-s_{ij}$ and the set of rates $\hat{k}_j$.
The reactants are the molecules of the two species. In our model the reactions are birth and death and 
the possible products are i) another molecule of the same species (birth), or  
ii) $\varnothing$ for the degradation of a molecule (death reaction).

\begin{figure}
\centering
\includegraphics[width=1.0\textwidth]{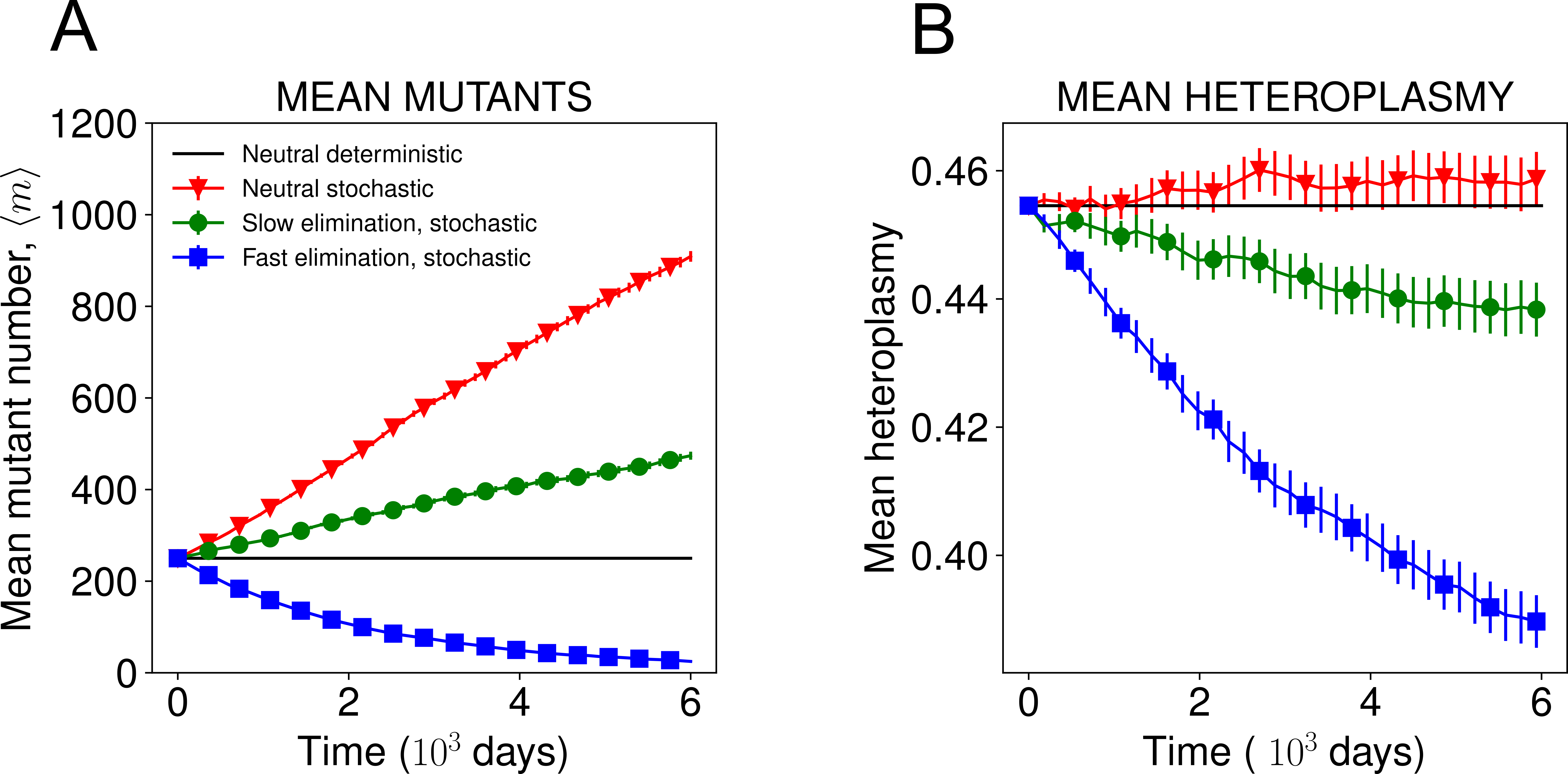}
\caption{(A) In the stochastic  one-region system, mean mutant copy number increases in the neutral model (red) and in presence of preferential eliminarion of mutants (green). This is a consequence of higher carrying capacity for mutants ($\delta<1$) and noise. In the
corresponding deterministic neutral model, mutant copy number remains constant (black). Parameter values in SI.\ref{toy_params}. Error bars are SEM. 
(B) In the one-region system, heteroplasmy is constant in the neutral deterministic system, as well as mean heteroplasmy in the neutral stochastic system (after a transient). Any selective elimination of mutants in the stochastic system causes a decrease in mean heteroplasmy, while  mean number of mutants can still increase (green line, see panel A). Parameter values in SI.\ref{toy_params}. Error bars are SEM. }
\label{fifth_fig}
\end{figure}

The chemical reaction network for the single-region model of the first section of Results is
\begin{equation}
\begin{split}
W& \myrightarrow{\mu} \varnothing \\
M& \myrightarrow{\mu} \varnothing \\
W& \myrightarrow{\lambda(w,m)} W+1 \\
M& \myrightarrow{\lambda(w,m)} M+1.
\end{split}
\label{stoch_model}
\end{equation}
This is a network with $N=2$ species and $R=4$ reactions, with stoichiometry matrix given by 
\begin{equation}
S=
\begin{bmatrix}
 -1& 0 & 1 & 0 \\
  0& -1 & 0& 1  
\end{bmatrix}
\end{equation}
The global rates for wildtypes or mutants are found by multiplying the
per molecule rates $\mu, \lambda$ by $w$ or $m$. This is true only in the case
$S_{ij} \leq 1$; more details for the general case can be found in Ref.~\cite{Aryaman19_net}. 
These reactions and rates can be used to set up a Chemical Master Equation (CME), 
a system of coupled ODEs in $P(w,m,t)$, the probability that 
the system is in the state $(w,m)$ at time $t$.
In principle, solving the CME would give the probability that the system is found in any state at any 
given time. However, given the nonlinearity of the global rates, the CME cannot be solved exactly.  One can explore the behaviour of the stochastic models by simulating the CME, which can be done  
through Gillespie's stochastic simulation algorithm (SSA) \cite{Gillespie76, Gillespie77}. This 
algorithm is exact, in that each Gillespie simulation represents an exact
sample from  $P(w,m,t)$.
The simulations allowed us to
observe the increase in mean mutant copy number $\langle m \rangle $ for $\delta <1$ (Fig.~\ref{fifth_fig}A, red line).

However, it is worthwhile having an analytical account of this effect, in order to know how 
the increase in mutant copy number depends on the parameters of the model.
We have obtained an effective description of the system, in the form of 
an approximate stochastic differential equation (SDE) that shows an increase in the 
number of mutants for $\delta <1$. The steps are the following:

\begin{enumerate}
\item Applying the Kramers-Moyal expansion to the CME, obtaining a Fokker-Planck equation  (a PDE for the probability distribution $P(w,m,t)$).

\item Converting the Fokker-Plack Equation into a system of two coupled SDEs in the variables
$w(t)$ and $m(t)$.

\item Applying a stochastic dimensionality reduction procedure, that exploits the fact that the system 
fluctuates around a central manifold (CM), to get a single  SDE for $m(t)$ that shows a positive drift
for $\delta <1$.
\end{enumerate}
The first two steps are standard \cite{Gardiner85, Jacobs10}. From the chemical reaction network in Eq.~\eqref{stoch_model}, we obtain the system of SDEs:
\begin{equation} 
\begin{split}
\d{} w&= cw(N_{ss}-w-\delta m) \d{} t +w[c(N_{ss}-w-\delta m)+2\mu]^{\nicefrac{1}{2}} \d{} W_1\\
\d{} m&= cm(N_{ss}-w-\delta m)\d{} t +m[c(N_{ss}-w-\delta m)+2\mu]^{\nicefrac{1}{2}}\d{} W_2, 
\end{split}
\label{eq:SDE_system}
\end{equation}
where $ \d{} W_1$ and $\d{} W_2$ are two i.i.d. Wiener increments, i.e. Gaussian noise with zero mean and variance $dt$ (see also Eq.~\ref{eq:Wiener}).
The third steps relies on a recently developed technique
\cite{Const16, Parsons17, Const17}, which 
works specifically for systems fluctuating around a CM. 
In the next section SI.\ref{maths} we detail each step, while here 
we only present and interpret the results. The final, effective SDE is
\begin{equation}
 \d{} m=\frac{2(1-\delta)\mu}{N_{ss}}  m\left ( 1-\frac{\delta m}{N_{ss}}\right ) \d{} t+ \frac{1}{N_{ss}}\left [2 m \mu (N_{ss}-\delta m)(N_{ss}+m-\delta m) \right ]^{\nicefrac{1}{2}} \d{}  W,
\label{mut_SDE}
\end{equation}
with $dW$ a  Wiener increment. The first term on the RHS
is the drift, and it corresponds to a logistic growth with carrying capacity 
$\frac{N_{ss}}{\delta}$. When the drift is positive, since $\langle dW \rangle =0$,
copy number increases on average.
The drift is positive for $0<\delta<1$, i.e. when mutants have a higher carrying capacity or density, and $\delta m < N_{ss} $.
Because of noise the final value of 
the mutant copy number will not be $\frac{N_{ss}}{\delta}$. The dynamics stop
either when $m=0$ or $m=\frac{N_{ss}}{\delta}$, the only values for which $dm=0$.
This approximation relies on the system fluctuating closely around the CM, which is true for 
large values of $c$ (strong control). The parameter $c$ is not present in Eq. (\ref{mut_SDE}) because 
this equation holds exactly for $c \rightarrow \infty$.
Indeed, when simulating Eq. (\ref{mut_SDE}) using the Euler scheme, 
the agreement with the results of the exact Gillespie simulations is 
better for larger values of $c$ (not shown).
By applying It\^{o}'s formula (see SI.\ref{maths}) it is possible to find the SDE for $h$, which is
\begin{equation}
\d{} h = \left [(2 \mu -cn+cN_{ss} + chn(1-\delta))\frac{ h(1-h)}{n} \right ]^\frac{1}{2} \d{} W,
\label{het_SDE}
\end{equation}
where $n=w+m$. We notice in this equation the absence of a drift term, meaning that
\begin{equation}
\frac{\d{} \langle h \rangle }{\d{} t} = 0,
\end{equation}
i.e. mean heteroplasmy is constant as stated in the main text. This is shown in Fig.~\ref{fifth_fig}B, that reports data from an ensemble of stochastic simulations (apart from the black line that refers to the ODE system). We see that the red line stabilises to a constant value after a transient. The transient is due to a difference in the steady state value of the effective population $w+\delta m$ between the deterministic and stochastic systems. The two systems are both initialised in the steady state of the deterministic one, hence the stochastic system first hits its steady state and then evolves as described.

A heuristic way to understand the increase in $\langle m \rangle$ is the following.
The per capita rates in Eq. (\ref{stoch_model}) are linear, hence the system is a  stochastic  
Lotka-Volterra model, for which each individual has the same chance of generating a lineage that will take over the whole system \cite{Czuppon18}. Hence, this probability must be $\frac{1}{w_0+w_0}$. Since there are $m_0$ mutants, the probability that a mutant will take over the whole population is $\frac{m_0}{w_0+w_0}=h_0$. The steady state population of mutants is $\frac{N_{ss}}{\delta}$. Based on this, one can conclude that the \textit{mean} number of mutants
$\langle m \rangle \rightarrow h_0 \frac{N_{ss}}{\delta}$ as $t \rightarrow \infty$. Taking the limit $t \rightarrow \infty$ means that we wait long enough for fixation to happen and for the population to reach the steady state. Recall that this differs from the deterministic case, in which there is a locus of fixed points of equation $w+\delta m = N_{ss}$ (the CM) and the system equilibrates at the point of the CM for which heteroplasmy is~$h_0$. For $0<\delta <1$, $h_0 \frac{N_{ss}}{\delta} > m_0$ always, hence
$\langle m \rangle$ will increase. 
It is also illuminating to consider the case in which $\delta \approx 0$, namely when the mutants 
are scarcely controlled. When a wildtype dies at steady state this increases the birth rate. 
Either a mutant or a wildtype will be the next birth, but if a mutant is born this leaves 
the elevated birth rate (caused by the degradation of the wildtype) almost unchanged,
encouraging further possible mutant births.

Interestingly, the stochastic system can exhibit an increase in $ \langle m \rangle$ even if 
we introduce preferential elimination of mutants (Fig.~\ref{fifth_fig}A, green line). This can be done by increasing mutant
degradation rate by  $\epsilon$, reflecting what we expect for defective mtDNA molecules. Indeed, under certain cases there is evidence for enhanced clearing of defective mutations \cite{Wei19, Shtolz19, Krishnan08}, subject to higher mitophagy rates. Mathematically, this is described by replacing the second reaction in  
Eq. (\ref{stoch_model}) by
\begin{equation}
M \myrightarrow{\mu + \epsilon} \varnothing,
\label{sel_dis}
\end{equation}
with $\epsilon>0$, leaving all other reactions unchanged.
In the next section (Eq.~\eqref{eq:general_effective_eps}) we will see that, through the techniques in Refs. \cite{Const16, Parsons17, Const17}, it is possible to derive an
SDE for mutant copy number that is valid for $\epsilon \ll \mu: $
 \begin{equation}
\d{} m=\left [\frac{2(1-\delta)\mu}{N_{ss}} -\epsilon \right ]  m\left ( 1-\frac{\delta m}{N_{ss}}\right ) \d{}t+ \frac{1}{N_{ss}}\left [2 m \mu (N_{ss}-\delta m)(N_{ss}+m-\delta m) \right ]^{\nicefrac{1}{2}} \d{}W.
\label{full_SDE}
\end{equation}
The drift in $m$ is now positive for $\delta m  < N_{ss}$ and $\epsilon<\frac{2(1-\delta)\mu}{N_{ss}}$. In this case,
the combined effect of stochasticity and differences in carrying capacity (density) produces 
a stochastic reversal of the direction of deterministic selection \cite{Const16}. Indeed, in the deterministic case, for any $\epsilon >0$ and for any initial condition with non-zero $w_0$ the system tends to the point $(N_{ss}, 0),$ the only stable fixed point, corresponding to wildtype fixation (mutant extinction).  Conversely, in the stochastic case, there is a critical  $\epsilon_M= \frac{2(1-\delta)\mu}{N_{ss}}$ such that, for $\epsilon < \epsilon_M$, $\langle m \rangle$ still increases, meaning that mutant fixation is more likely than wildtype.
Inspection of the expression for $\epsilon_M$, interpretable as the largest amount of additional mitophagy rate that still allows mutant to expand, reveals the following  interesting points:
 \begin{itemize}
 
  \item  $\epsilon_M$ increases with $\mu$, meaning that the noisier the system, the higher the mitophagy rate that mutants can tolerate and still expand.
 
 \item For $0<\delta <1$, $\epsilon_M$ is decreasing function of $\delta$. The higher the the difference in density, the higher the mitophagy rate that mutants can overcome. 
 
 \item $\epsilon_M$ is a decreasing function of $N_{ss}$. In particular, $\epsilon_M \rightarrow 0$ when
 $N_{ss} \rightarrow \infty$. This is common to stochastic effects, whose intensities decrease with the system size, vanishing in the deterministic limit for which system size tends to infinity.
 \end{itemize}

\section{Derivation of the effective SDE for the stochastic model constrained to the CM}
\label{maths}

In this section we show that Eq.~\eqref{mut_SDE} is an effective, approximate description
for the system defined by the chemical reaction network in  Eq.~\eqref{stoch_model}
given the replication rate in Eq.~\eqref{linear_f_c_SI}. We provide detail
for each of the steps listed in SI.\ref{1_region_SI}.
The approach used here is adapted from standard texts \cite{Gardiner85, Jacobs10} and from
Refs. \cite{Parsons17, Const17}.
We strive to provide a self-contained treatment, hoping to make our derivation and the set of
techniques accessible to a wider audience. Our exposition partly follows Ref. \cite{Aryaman19_net},
where the same procedure is used for a related problem.

\paragraph*{1 - From the CME to a Fokker-Planck equation via Kramers-Moyal expansion. \\}

Let us denote $\mathbf{n} = (w,m)$, a vector
specifying the system state. The CME can be written in compact form as
\begin{equation}
\deriv{P(\mathbf{n},t)}{t}= \sum_{\mathbf{n'} \neq \mathbf{n}}\left[T(\mathbf{n}|\mathbf{n}')P(\mathbf{n}',t)-T(\mathbf{n}'|\mathbf{n})P(\mathbf{n},t)\right]\label{eq:master_eqn_continuous_omega}
\end{equation}
$T(\mathbf{n}|\mathbf{n}')$ is the transition rate from state 
$\mathbf{n'}$ to $\mathbf{n}$.  Given the reaction network in Eq.~\eqref{stoch_model},
the corresponding (global) transition rates are 
\begin{equation}
\begin{split}
T&(w-1,m | w,m) = w \mu \\
T&(w,m -1 | w,m) = m \mu \\
T&(w+1,m | w,m) = w \lambda(w,m) \\
T&(w,m+1 | w,m) = m \lambda(w,m)
\end{split}
\label{eq:trans_rates}
\end{equation}

The first step is to write the CME in a continuous form. We replace copy numbers $\mathbf{n}$
by the  abundances $\mathbf{x} = \frac{\mathbf{n}}{N_{ss}}$ relative to the effective
population size.
The variable $\mathbf{x}$ can be considered continuous for large $N_{ss}$. 
The CME becomes
\begin{equation}
\pderiv{P(\x,t)}{t}= \int_{-\infty}^{\infty}\d{}\x{}'\left[T(\x{}|\x{}')P(\x{}',t)-T(\x{}'|\x{})P(\x{},t)\right].
\end{equation}
We now proceed by expanding the CME through the second-order multivariate Kramers-Moyal expansion \cite{Gardiner85},
that can be written as 
\begin{equation}
\pderiv{P(\x,t)}{t} \approx \int_{-\infty}^{\infty}\left(-\nabla \left(T(\x'|\x)P(\x,t)\right)^T \cdot (\x'-\x) + \frac{1}{2} (\x'-\x)^T \cdot \mathbf{H}_{\x'}(\x) \cdot (\x'-\x) \right)\d{}\x' \label{eq:KM_mv}
\end{equation}
where $\mathbf{H}_{\x'}(\x)$ is the Hessian matrix of $T(\x'|\x)P(\x)$, whose general expression is
\begin{equation}
\mathbf{H}_{\x'}(\x)\defeq\begin{pmatrix}
\frac{\partial^2}{\partial x_1^2}  & \dots & \frac{\partial^2}{\partial x_1 \partial x_N}  \\
\vdots & & \vdots \\
\frac{\partial^2}{\partial x_N \partial x_1}  & \dots & \frac{\partial^2}{\partial x_N^2} 
\end{pmatrix} T(\x'|\x)P(\x).
\label{eq:Hessian}
\end{equation}
A transition $\x \rightarrow \x'$ corresponds to some reaction $j$ which moves the state from $\x$ to 
$\x'$. Since we know how each reaction affects state $\x$ through the constant 
stoichiometry matrix $S_{ij}$, and since the global rate of each reaction is independent 
from $\x'$ itself (see Eq.~\eqref{eq:trans_rates}), we may transition from a notation 
involving $\x$ and $\x'$ into a notation involving $\x$ and the reaction $j$ that affects $\x$ . 
We may therefore define $T_j(\x) \defeq T(\x'|\x)$, 
and let $\mathbf{H}_{\x'}(\x) \rightarrow \mathbf{H}_j(\x)$.

We now wish to re-write Eq.~\eqref{eq:KM_mv} as a Fokker-Planck equation. Since the integral in 
Eq.~\eqref{eq:KM_mv} is over $\x'$, and for a given $\x$ every $\x'$ corresponds to a reaction $j$, we may interpret the
integral in Eq.~\eqref{eq:KM_mv} as a sum over all reactions, 
i.e.\ $\int \d{}\x' \rightarrow \sum_{j=1}^R$. Hence, for the $j$\textsuperscript{th} 
reaction, $[(\x'-\x)]_i = S_{ij}$. With these observations, 
we may write the first term in the integral in Eq.~\eqref{eq:KM_mv} as
\begin{align}
\int_{-\infty}^{\infty}-\nabla (T(\x'|\x)P(\x,t))^T \cdot (\x'-\x) \d{}\x' &= \int_{-\infty}^{\infty}-\nabla (T_j(\x)P(\x,t))^T \cdot (\x'-\x) \d{}\x' \nonumber \\
&= -\sum_{j=1}^R \sum_{i=1}^N \pderiv{}{x_i}( T_j(\x) P(\x,t))  S_{ij} \nonumber \\
&= - \sum_{i=1}^N \pderiv{}{x_i} A_i P(\x,t)
\end{align}
where
\begin{equation}
\mathbf{A} \defeq \mathbf{ST},
\end{equation}
with $\mathbf{A}$ a vector of length $N$, $S$ the $N\times R$ stoichiometry matrix and $\mathbf{T}$ 
the vector of transition rates, of length $R$. 
To re-write the second integral of Eq.~\eqref{eq:KM_mv}, we write an element of 
the Hessian $\mathbf{H}_j$ in Eq.~\eqref{eq:Hessian} as
\begin{equation}
H_{jlm} = \frac{\partial^2}{\partial x_l \partial x_m} T_j(\x) P(\x,t)
\end{equation}
where $j=1,\dots,R$ and $l,m=1,\dots,N$. Thus, we may write
\begin{align}
\int_{-\infty}^{\infty} \frac{1}{2} (\x'-\x)^T \cdot \mathbf{H}_{\x'}(\x) \cdot (\x'-\x) \d{}\x' &= \frac{1}{2} \sum_{j=1}^R \sum_{l=1}^N \sum_{m=1}^N S_{lj} H_{jlm} S_{mj} \nonumber \\
&= \frac{1}{2} \sum_{j=1}^R \sum_{l=1}^N \sum_{m=1}^N S_{lj} \frac{\partial^2}{\partial x_l \partial x_m} T_j P(\x,t) S_{mj} \nonumber \\
&= \frac{1}{2} \sum_{l=1}^N \sum_{m=1}^N  \frac{\partial^2}{\partial x_l \partial x_m} \left( \sum_{j=1}^R S_{lj} T_j S_{mj} \right) P(\x,t) \nonumber \\
&= \frac{1}{2} \sum_{i,m=1}^N   \frac{\partial^2}{\partial x_i \partial x_m} B_{im} P(\x,t)
\end{align}
where
\begin{equation}
\mathbf{B} \defeq \mathbf{S} \cdot \Diag(\mathbf{T}) \cdot \mathbf{S}^T, \label{eq:diff_KM_fpe}
\end{equation}
and $\mathbf{B}$ is an $N\times N$ matrix, and $\Diag(\mathbf{T})$ is a diagonal matrix whose main diagonal is the vector $\mathbf{T}$. We may therefore re-write Eq.~\eqref{eq:KM_mv} as a Fokker-Planck equation for the state vector $\x$ of the form
\begin{equation}
\pderiv{P(\x,t)}{t} \approx - \sum_{i=1}^N \pderiv{}{x_i} [A_i(\x)P(\x,t)]+\frac{1}{2}\sum_{i,m=1}^N \pdd{}{x_i}{x_m}[B_{im}(\x)P(\x)]. 
\label{eq:FPE}
\end{equation}
From a system of coupled ODEs for $P(w,m,t)$  (the  CME of Eq.~\eqref{eq:master_eqn_continuous_omega})  we have obtained a single PDE for $P(\x,t)$ approximating 
copy numbers as continuous variables.

\paragraph*{2 - From Fokker-Planck to a system of It\^{o} SDEs. \\}

In general, the Fokker-Planck equation in Eq.~\eqref{eq:FPE} is equivalent \cite{Jacobs10} to the following It\^{o} stochastic differential equation (SDE)
\begin{equation}
\d{} \x = \mathbf{A} \d{} t + \sqrt{\sigma}\mathbf{G} \d{} \mathbf{W} 
\label{eq:SDE}
\end{equation}
where $\mathbf{G}\mathbf{G}^T\equiv\mathbf{B}$ (where $\mathbf{G}$ is an $N \times R$ matrix)  and $\d{} \mathbf{W}$ is a vector of length $R$ of independent Wiener increments, that satisfy
\begin{equation}
\int_0^t \d W \defeq W(t),\ P(W,t) \equiv \frac{1}{\sqrt{2 \pi t}} e^{-W^2/(2t)},
\label{eq:Wiener}
\end{equation}
and $\sigma$ is a constant that controls the strength of the noise. For our chemical reaction network Eq.~\eqref{stoch_model}, the corresponding system of It\^{o} SDEs is given in Eq.~\eqref{eq:SDE_system}.

\paragraph*{3 - From a system of It\^{o} SDEs to a single effective SDE for a system
forced onto the central manifold. \\}

What follows is an adaptation from \cite{Const17, Parsons17}, to which we refer for a proof and 
for interpretation of the functions introduced.
We start from Eq.~\eqref{eq:SDE}. The procedure can be applied when the system admits a manifold
on which $\mathbf{A}=0$, the central manifold (CM). We though it would be an helpful complement to Refs. \cite{Const17, Parsons17} to give an explicit sequence of the steps involved for the specific system at hand.

\begin{itemize}

\item Identify the CM.

\item Find the Jacobian of $\mathbf{A}$, namely $J$ such that 
$$J_{ij} =\pderiv{A_i}{x_j},$$
and evaluate it on the CM, obtaining $J_{CM}$.

\item Find the eigenvalues $\lambda_1, \cdots, \lambda_N$  of  $J_{CM}$.
The CM is the kernel of $J_{CM}$, hence the multiplicity of the zero eigenvalue is the dimensionality of the CM.

\item Compute the decomposition of $J_{CM}$,
writing 
$$ J_{CM} = W \Lambda W^ {-1},$$
where $W = (\mathbf{w_1}, \cdots, \mathbf{w_N})$ is the matrix of eigenvectors with those corresponding 
to the zero eigenvalues written first.

\item Compute $J_{CM}^+$, defined as  
$$ J_{CM}^+ = W^{-1} \Lambda^+ W,$$
where $\Lambda^+$ is the diagonal matrix with diagonal elements $\lambda_1^+, \cdots, \lambda_N^+$
defined as
$$  \lambda_i^+ =
    \begin{cases}
      \nicefrac{1}{\lambda_i} & \text{if $\lambda_i \neq 0$}\\
      0 & \text{if $\lambda_i = 0.$}\\
    \end{cases}       
$$
\item For each element  of $\mathbf{A}$, compute the Hessian $H_i$ as

$$ H_{ijk} = \frac{\partial^2 A_i}{\partial x_i \partial x_j}, $$ 

and evaluate it on the CM, obtaining $H_{i_{CM}}.$

\item Compute the matrix $\mathcal{P}$ as
$$\mathcal{P}= I-J^+ J, $$ 
where $I$ is the identity matrix, and the matrices $Q_i$ as
$$Q_i=-\sum_{l=1}^N [J_{il}^+\mathcal{P}^T H_l \mathcal{P} + \mathcal{P}_{il}(J^{+T} H_l \mathcal{P}+\mathcal{P}^T H_lJ^+) ] $$ 

\item Calculate the vector  $\mathbf{g(\x)}$, whose elements are 
$$
g_i(\x)  = \frac{1}{2}\Tr[\mathbf{G}^T Q_i \mathbf{G}].
$$

\item Finally, effective SDEs for the variable $\mathbf{z}$, namely the variable 
$\x$ constrained to the CM, are  
\begin{equation}
\d{} \z = \sigma \mathbf{g(\z)} \d{} t + \sqrt{\sigma} \mathcal{P}\mathbf{G(\z)} \d{} \mathbf{W},
\label{eq:general_effective}
\end{equation} 
where the equations are now \textit{uncoupled} because all the functions are evaluated on 
the CM.
\end{itemize}
This procedure allows one to obtain Eq. \eqref{mut_SDE} as an effective
description of the neutral stochastic model defined in Eq. \eqref{stoch_model}. We provide a Mathematica notebook (see Supplementary File 1) to obtain Eq.~\eqref{eq:SDE} from the system of SDEs in Eq.~\eqref{eq:SDE_system}. 
This technique is extended in Ref. \cite{Parsons17} to the more general case in which the stochastic 
system is equivalent to a system of It\^{o} SDEs of the form 
\begin{equation}
\d{} \x = ( \mathbf{A} +\epsilon \mathbf{h}) \d{} t + \sqrt{\sigma} \mathbf{G} \d{} \mathbf{W},
\end{equation}
where $\mathbf{h} \equiv \mathbf{h(x)}$ is another vector-valued function of $\x$ and  $\epsilon \ll \sigma $ is a small parameter,
such that for $\epsilon=0$ the system admits a CM. In this case, Eq.~\eqref{eq:general_effective}
becomes
\begin{equation}
\d{} \z = \left [\epsilon \mathcal{P} \mathbf{h (\z)} + \sigma \mathbf{g(\z)} \right ] \d{} t + \sqrt{\sigma} \mathcal{P}\mathbf{G(\z)} \d{} \mathbf{W}.
\label{eq:general_effective_eps}
\end{equation} 
The stochastic system in which mutants are subject to preferential elimination $\epsilon$  (Eq.~\eqref{sel_dis})
corresponds to the case $\mathbf{h(x)}=(0,x_2)$ (recall that $x_2 = \nicefrac{m}{N_{ss}}$).
Inserting this into Eq.~\eqref{eq:general_effective_eps}, one obtains Eq.~\eqref{full_SDE}, that describes the noise and density-induced selection reversal.

\paragraph*{Change of variable through It\^{o}'s formula to obtain an SDE for heteroplasmy. \\}
It\^{o}'s formula states that, for an arbitrary function $y(\x,t)$ where $\x$ satisfies Eq.~\eqref{eq:SDE}, we may write the following SDE:
\begin{equation}
\d y(\x,t) = \left\{ \pderiv{y}{t} + \left(\nabla y\right)^T \mathbf{A} + \frac{1}{2} \Tr \left[ \mathbf{G}^T \mathbf{H'}(\x) \mathbf{G} \right]   \right\} \d t + (\nabla y)^T \mathbf{G} \d{} \mathbf{W}, \label{eq:ito_formula}
\end{equation}
where $\mathbf{H'}(\x)$ is the Hessian matrix of $y(\x,t)$ with respect to $x$
(see Eq.~\eqref{eq:Hessian}, where $T(\x'|\x) P(\x)$ should be replaced with $y(\x,t)$). 
Applying this rule to Eq.~\eqref{mut_SDE} for 
$$
y(\x,t)=(h(\x),n (\x)),
$$
where $h$ is the heteroplasmy and $n$ is the total copy number,
one obtains Eq.~\eqref{het_SDE}
\section{The two-region model: deterministic and stochastic formulation}
\label{2_regions_SI}

The deterministic two-region model is formalised via the ODE system
\begin{equation} 
\begin{split}
\dot{w_1}&= cw_1[N_{ss}-(w_1+\delta m_1)]+\gamma (w_2-w_1) \\
\dot{m_1}&= cm_1[N_{ss}-(w_1+\delta m_1)]+\gamma (m_2-m_1) \\
\dot{w_2}&= cw_2[N_{ss}-(w_2+\delta m_2)]+\gamma (w_1-w_2) \\
\dot{m_2}&= cm_2[N_{ss}-(w_2+\delta m_2)]+\gamma (m_1-m_2). 
\end{split}
\label{ODE_system_2}
\end{equation}
The addition of the second term on the RHS,
proportional to the constant hopping rate $\gamma$ 
accounts for diffusion of the mtDNA molecules between neighbouring regions.
The first term on the RHS of each equation means that each cell seeks to maintain 
its own separate target population, i.e. copy number control is local.
This system does not admit an analytical solution. However, it has a CM given by 
$w_1 + \delta m_1=w_2 + \delta m_2=N_{ss}$, $w_1=w_2$ (and consequently $m_1=m_2$). 
It is easy to see that if these conditions are satisfied,
all the time derivatives (the left-hand-sides) are zero and the dynamics stop, 
i.e. steady state is reached. A similar system of ODEs can be used to describe chains of regions of
arbitrary length (see SI.\ref{pheno_wave_speed}).

The stochastic formulation of the model is obtained by replicating the reaction network 
in Eq. (\ref{stoch_model}) for $w_2,m_2$ and adding the reactions accounting for diffusion of the 
mtDNA molecules. The death rate is still constant and common, and each region has its own independently controlled replication rate given by $\lambda_i=\mu+c(N_{ss}-w_i-\delta m_i)$, for $i=1,2$. The full chemical reaction network is
\begin{equation}
\begin{split}
W_1& \myrightarrow{\mu_1} \varnothing \\
M_1& \myrightarrow{\mu_1} \varnothing \\
W_1& \myrightarrow{\lambda_1(w_1,m_1)} W_1+1 \\
M_1& \myrightarrow{\lambda_1(w_1,m_1)} M_1+1 \\
W_2& \myrightarrow{\mu_2} \varnothing \\
M_2& \myrightarrow{\mu_2} \varnothing \\
W_2& \myrightarrow{\lambda_2(w_2,m_2)} W_2+1 \\
M_2& \myrightarrow{\lambda_2(w_2,m_2)} W_2+1 \\
W_1& \myrightarrow{\gamma} W_2 \\
M_1& \myrightarrow{\gamma} M_2 \\
W_2& \myrightarrow{\gamma} W_1 \\
M_2& \myrightarrow{\gamma} M_1. \\
\end{split}
\label{stoch_model_2}
\end{equation}
The dimensionality reduction technique from \cite{Parsons17}, used to obtain Eq. (\ref{full_SDE}) is not 
viable in this case to obtain an SDE showing the increase in $\langle h_1 \rangle $ 
(or $\langle h_2 \rangle $ ) when $0 <\delta <1$. The technique works by projecting the system onto the CM,
where the two regions are identical; however, the increase in mean heteroplasmy is driven by fluctuations 
that move one of the regions away from the CM condition, after which the system relaxes to steady state with
an average increase in $h$. The approximation involved in the technique is too crude to
capture the effect. However, stochastic simulations consistently show the increase 
in $\langle h \rangle$
in chains of arbitrary length when $0 <\delta <1$. This effect, conveniently 
highlighted in the two-region system (see Fig.~\ref{first_fig}D), is key to the 
appearance of a wave of heteroplasmy in a chain of regions.

\section{The one-region model is a microscopic building block for a system exhibiting a noise-driven wave}
\label{chain_SI}
In Ref. \cite{Hallat11}, a stochastic continuous-space  model is presented, in the form
of partial differential equations, describing a noise-driven wave of mutants, possibly subject to higher death rates than wildtypes.
The paper is an extension of the classical Fisher-Kolmogorov PDE to a stochastic setting.
In this section, we show how that our microscopically interpretable bottom-up model, 
based on the simple Lotka-Volterra dynamics,
gives rise to the model in  Ref. \cite{Hallat11} for $\delta \rightarrow 1 $ and in a continuous-space limit.

Let us reconsider the deterministic system of Eq. (\ref{ODE_system}) and change variable to 
$(n,h)$, where $n = w+m$ is the total population. We already observed that $\frac{\d{} h}{\d{} t}=0$.
For $n$ we have
\begin{equation}
\frac{\d{}n}{ \d{}t}=cn[N_{ss}-w-\delta m]
\end{equation}
Let's write $\delta=1-\alpha$ :
\begin{equation}
\frac{\d{}n}{\d{}t}=cn[N_{ss}-w-m+\alpha m]
\end{equation}
The constancy of $h$ allows us to write $m=hn$, hence
\begin{equation}
\frac{\d{}n}{\d{}t}=cn[N_{ss}-n(1-\alpha h)] = cN_{ss} n \left [1-\frac{n}{N_{ss}/(1-\alpha h)} \right].
\end{equation}
Considering the case $\alpha \rightarrow 0$, equivalent to $\delta \rightarrow 1$, we can write  
$$\frac{1}{1-\alpha h} =1+\alpha h + \mathcal{O}(\alpha^2).$$
Neglecting $\mathcal{O}(\alpha^2)$ terms we obtain 
\begin{equation}
\frac{\d{}n}{\d{}t}=cN_{ss} n \left [1-\frac{n}{K(h)} \right ],
\end{equation} 
with $K(h)=N_{ss}(1+\alpha h)$. The total copy number follows a logistic 
growth  with a carrying capacity $K(h)$ that is 
a linearly increasing function of heteroplasmy $h$. 
In order to model a spatially-extended system as a muscle fibre, we  give spatial dependence 
to copy number, i.e. $n \equiv n(x)$. The hopping of mtDNA molecules between neighbouring regions 
of the muscle fibres is described in this case as a diffusion term, proportional to the second spatial
derivative of $\frac{\partial ^2 n(x)}{\partial x^2}$. In order to model a muscle fibre, we can therefore write
\begin{equation}
\frac{\partial n(x,t)}{\partial t}= D \frac{\partial ^2 n(x,t)}{\partial x^2}+ cN_{ss} n(x,t) \left[1-\frac{n(x,t)}{K(h(x,t))} \right],
\label{start_Hall}
\end{equation} 
where we have explicitly written the time-dependence.  $D$ is the diffusion coefficient, for which
$D=\gamma L^2$, where $L$ is the inter-region spacing.

Eq. (\ref{start_Hall}), derived from the special case of the simple single-region Lotka-Volterra model, 
is the starting point of the work in Ref. \cite{Hallat11}. By introducing noise from 
Wright-Fisher sampling and taking the low-diffusion limit $D \rightarrow 0$, a wave equation for $h$ can be derived which holds also in the case of slow preferential elimination of mutants. 
The limit $\delta \rightarrow 1$, that we have taken here to derive Eq. (\ref{start_Hall}) from 
Eq. (\ref{ODE_system}), is needed also in Ref. \cite{Hallat11} in order to derive the wave equation. 
Hence our model, for some limiting 
values of the parameters $D$ and $\delta$, leads to a wave-equation for $h$. We stress the remarkable fact that from one of the simplest models of population genetics, 
the Lotka-Volterra,
it is possible to derive an equation for a travelling wave of heteroplasmy driven 
by stochastic survival of the densest, albeit in some limit.

\begin{figure}
\centering
\includegraphics[width=0.95\textwidth]{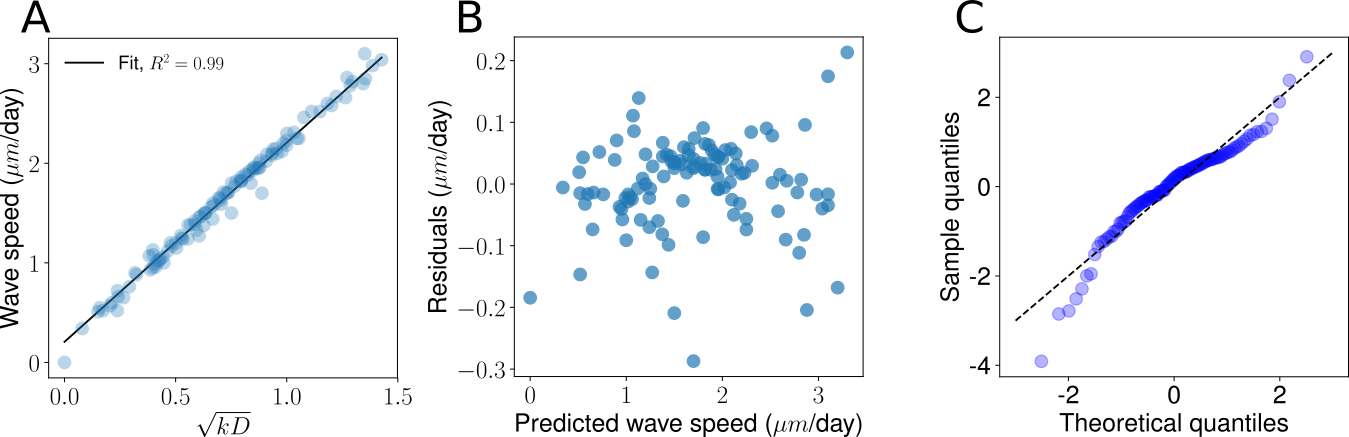}
\caption{An empirical formula accurately predicts 
the  survival of the densest wave speed obtained through stochastic simulations for a wide range of 
parameter values. A) The blue dots are 110 values of wave speed measured through stochastic simulations
of the linear feedback control for varying values of the parameters $N_{ss}, \mu, \gamma$ and $0<\delta<1$.
The black line is the linear fit against the variable
$x~=~(1-\delta) \nicefrac{(\mu \gamma^3)^\frac{1}{4}}{N_{ss}^\frac{1}{3}}$. The excellent
fit $(R^2=0.99)$ shows that the wave speed $v$ is well predicted by Eq.~\eqref{eq:empirical_wavespeed}, except
for $x \rightarrow 0$. B)Although the residuals do not follow
a Gaussian distribution (Shapiro-Wilk, $p< 10^{-4}$), the absence of a clear pattern in the plot suggests that Eq.~\eqref{eq:empirical_wavespeed} is a satisfactory approximation in the parameter range considered. C) The Q-Q plot of the residuals against a normal distribution confirms the absence of a clear pattern.  }
\label{sixth_fig}
\end{figure}

\section{Phenomenological formula for wave speed}
\label{pheno_wave_speed}

In the previous section we have shown how our model, in the limits $\gamma \rightarrow 0$ and $\delta \rightarrow 1$, leads to an analytical description of 
a noise-driven wave. Moreover, through numerical simulations we have shown that the wave-like 
expansion is exhibited by our model without having to assume the above limits. 
We have performed simulations of a 201-region chain with   
110 combination of the parameter values,
with $\delta \in [0.1, 1]$, $N_{ss} \in [2,35]$, $\gamma \in [0, 0.25]$, $\mu \in [0.01, 0.15]$. For each parameter configuration we have observed a wave-like expansion, except, as expected, when $\delta=1$ or $\gamma=0$.
We have calculated the wave speed plotting 
the wavefront at different times and measuring the distance covered, measuring the degree of shift between the two wavefronts. In some 
cases, the overlap could not be total since the wavefront's steepness changed slightly over time: in those cases we have looked for an overlap of the midpoint of the curves (i.e. the point for which $ \langle h \rangle =0.5$).
In all cases, it has been verified that the speed is constant, by measuring it for several different time intervals.
We have found that
$v$ can be predicted with excellent accuracy through the formula
\begin{equation}
v=  2 \sqrt{kD} +\beta.
\label{eq:empirical_wavespeed}
\end{equation}
Apart from the intercept, Eq.~\eqref{eq:empirical_wavespeed} is analogous to the formula for the wave speed of FK waves, 
with $k$ being an effective 
reaction rate induced by stochastic survival of the densest, given by 
\begin{equation}
k_{SSD}=\frac{\sqrt{(1-\delta)^2 \mu \gamma}}{N_{ss}^\frac{2}{3}}, 
\label{eq:effective_reac_rate}
\end{equation} 
and $\beta= 0.20723 \pm 4 \cdot 10^{-5}$ is calculated via MLE. 
In Fig.~\ref{sixth_fig}A we show the fit, for which $R^2=0.99$, and we observe
residuals without any clear pattern. However, the presence of an intercept $\beta \neq 0$ means that Eq.~\eqref{eq:empirical_wavespeed} is not accurate for $x \rightarrow 0$, for which $v=0$. In Fig.~\ref{fourth_fig}A we compare the probability distribution of the wave speed predicted through Eq.~\eqref{eq:empirical_wavespeed}, obtained through Monte-Carlo sampling on the basis of the estimated distributions for the parameters of our model, given in SI.\ref{parameters}.

Finally,  we have verified that in a deterministic model deployed on a chain, i.e.  Eq.~(\ref{ODE_system_2}) applied  to a larger number of regions, the 
high-heteroplasmy front diffuses away even if $\delta <1$ (Fig.~\ref{seventh_fig}A), which is consistent with the statement that the wave-like expansion is noise-driven.

\section{Further links to the wider literature}
\label{our_differences}
Density-dependent selection refers to situations in which the fitnesses of the  different species within a population depend differently on population density (or size) \cite{Lande09}. The theory of \textit{r/K} selection \cite{Clarke72, MacArthur67}, one of the most influential ideas in evolutionary biology \cite{Pianka70, Boyce84, Reznick02}, examines how selection shapes the strategies of species in the presence (or in absence) of density effects. The study of density-dependent selection has attracted interest across five decades \cite{Clarke72, Mueller97, Farkas14, Bertram19}, including in systems where stochasticity is important \cite{Slatkin79, Turelli80, Heckel80, Lande09, Parsons07I, Parsons07II, Oizumi16}.
In stochastic survival of the densest, wildtypes and mutants have a  common per capita replication rate. Per capita degradation rates are constant, and the mutant degradation rate can be higher than the wildtype by a constant additive factor. The mutant and wildtype rates therefore depend in the same way on copy numbers, and hence our model is not an example of density-dependent selection. Frequency-dependent selection \cite{Ayala74} is distinct, but related, to density-dependent selection. Numerous definitions of frequency dependent-selection have been used over the decades, a classic one being that the relative fitness of a species varies with the relative frequencies of other species \cite{Heino98}. Our model satisfies a strong mathematical definition of frequency-\textit{independent} selection, based on the per capita growth rate (PCGR) of the species, the difference between per capita birth and death rates. The PCGR can be defined as a vector-valued function of copy number variables, with an output for each species. Frequency-independent selection is assured when the Jacobian of the PCGR is independent of copy numbers \cite{Debenedictis77}. This is clear for our linear feedback model (and all generalised Lotka-Volterra models), since the replication rate in Eq.~\eqref{linear_f_c} is linear in $w$ and $m$. Therefore, stochastic survival of the densest is not a case of frequency-dependent selection.

In the context of a single-species population, the Allee effect can be described as a positive dependence of the PCGR on the density or size of the population for some interval of the values of population size \cite{Allee27}. This contrasts with theories in which the PCGR always decreases with population size because of competition for resources or space among members of the same species (intra-specific competition), as in logistic growth. When the dynamics of a population exhibits the Allee effect, it means that there is some other mechanism at play that counterbalances the general limiting effect of competition on the PCGR. An example is that of plants that reproduce through pollination, whose rate is increased by the density of plants in the environments. The Allee effect is connected to altruism, in that the efficiency of some fitness-increasing behaviours, e.g. defence and feeding, is enhanced by a large population size when the associated tasks are carried out cooperatively \cite{Courchamp99}. The difference between stochastic survival of the densest and the Allee effect is two-fold.
The Allee effect refers to the possibility that PCGR in a population can be an increasing function of population size (for some interval) despite the increase in intra-specific competition that generally comes with higher density. Our work instead models inter-specific (between species) competitions, showing that the species that is \textit{denser in isolation} can prevail despite a higher degradation rate and no replicative advantage. Moreover, the PCGR of our neutral model  is proportional to $N_{ss}-w-\delta m$ for both species; in the case of a higher degradation rate for mutants, their PCGR is proportional to $N_{ss}-w-\delta m$. In both cases we have linearly decreasing functions of the population size, with no positive dependence on it.

Previous work has explored how the spatial structure of a population can favour \cite{Nowak92, Lion08}  or  hinder \cite{Hauert04} the evolution of altruistic behaviour. These studies refer to game theoretical models that can be recast in terms of population dynamics models, due to the correspondence between the replicator equation and the generalised Lotka-Volterra model, in their deterministic \cite{Hofbauer98} and stochastic \cite{Constable17} versions. Similar results have been found applying agent-based models to cell populations \cite{Kreft04, Nadell10}.
In these studies, the \textit{continuous} spatial structure of the population influences the way its members interact, with interaction possible only within spatial neighbourhoods. The recurring theme of these studies is that when altruists cluster together they share the benefits of their behaviour mainly among themselves, and this can favour the spread of altruism.
Our work is different in that we have a \textit{discrete} spatial  structure: individuals can migrate to different regions, but the population itself within a region is well-mixed; all its members interact in the same way with one another and everyone benefits from the mutants' altruism in the same way. In this sense, our model is more similar to Refs. \cite{Const16, Houch12, Houch14}.

\section{ Parameter values for simulations in Figs.~\ref{first_fig} and \ref{fifth_fig} } 
\label{toy_params}
In this section we specify the simulations and list the parameters for  Figs.~\ref{first_fig}D and \ref{fifth_fig}, with the aim of allowing a smooth reproduction of the results. The parameter values are not derived from the scientific literature and are not meant to describe a realistic system. Rather, they are chosen to show the increase in mean mutant copy number and the increase in mean heteroplasmy with a limited computational effort in the simplest possible systems. Realistic parameter values for simulating skeletal muscle fibres models as a chain of regions (Fig.~\ref{second_fig}) are reported in the next section.

Fig.~\ref{first_fig}D refers to the two-region system. Data obtained simulating the deterministic ODE system Eq.~\eqref{ODE_system_2} (black line) and the chemical reaction network Eq.~\eqref{stoch_model_2} (other lines) with parameters $\gamma=\mu=7 \cdot 10^{-2}$/day, $\delta=10^{-1}$, $c=10^{-3}$/day, $N_{ss}=455.$ 
The chemical reaction network is simulated via the Gillespie algorithm. One region is initialised with zero heteroplasmy $w_{10}=455,m_{10}=0$ and the other with $w_{20}=450,m_{20}=50$, corresponding to an initial heteroplasmy of $10^{-1}$. 
For the neutral and deterministic cases mutants and wildtypes have the same degradation rate $\mu$. For the cases of slow and fast preferential elimination the degradation rate of mutants is increased by an additive constant $\epsilon$ (see Eq.~\eqref{sel_dis}). The increase in degradation rate is $\epsilon=1.5\cdot 10^{-3}$/day for slow elimination and $\epsilon=6.25 \cdot 10^{-3}$/day for fast elimination. For the deterministic system the  quantity plotted is $h_{mean}=(h_1+h_2)/2$, the mean of the heteroplasmies of the two regions. For the stochastic system, we plot  $\langle h_{mean} \rangle =(\langle  h_1+h_2 \rangle)/2,$ averaging over an ensemble of $3.6 \cdot 10^4$ simulations. Error bars are standard error of the mean (SEM). Fewer ($10^3$) realizations are enough to highlight a statistically significant change in $\langle h_{mean} \rangle$ in the stochastic systems.

Fig.~\ref{fifth_fig} refers to the one-region system. 
Data is obtained simulating the deterministic ODE system Eq.~\eqref{ODE_system} (black lines) and the chemical reaction network Eq.~\eqref{stoch_model} via Gillespie algorithm (other lines) with parameters $\mu=7 \cdot 10^{-2}$/day, $\delta=10^{-1}$, $c=2.5 \cdot 10^{-3}$/day, $N_{ss}=325$. The system is initialised at steady state with  $w_{0}=300,m_{0}=250$ corresponding to an initial heteroplasmy $h_0 \approx 0.45$. For the neutral and deterministic cases mutants and wildtypes have the same degradation rate $\mu$. For the cases of slow and fast preferential elimination the degradation rate of mutants is increased by an additive constant $\epsilon$ (see Eq.~\eqref{sel_dis}. The increase in degradation rate is $\epsilon=2.935\cdot 10^{-4}$/day for slow elimination and $\epsilon=1.1735 \cdot 10^{-3}$/day for fast elimination. In panel A, for the deterministic system we plot mutant copy number $m$; for the stochastic systems we plot $\langle m \rangle$, averaging over an ensemble of $10^4$ simulations. In panel B, for the deterministic system we plot heteroplasmy $h$; for the stochastic systems we plot $\langle h \rangle$, averaging over an ensemble of $10^4$ simulations. Fewer ($10^3$) realizations are enough to highlight a statistically significant changes in $\langle h \rangle$ and $\langle m \rangle$.

\section{Estimates parameter values and setup of numerical simulations}
\label{parameters}

In this section, we report and justify our best estimates of the parameter values used to simulate the spread of mtDNA deletions in the skeletal muscle fibres of rhesus monkeys. Specifically, we discuss the point estimates used to obtain Figs.\ref{second_fig}D, E, F and the estimated ranges used to obtain Fig.~\ref{fourth_fig}A. We derive these values from the scientific literature, in order to simulate a realistic travelling wave of mutants and compare the wave speed predicted by our simulations with that estimated from experiments (Fig.~\ref{second_fig}C).

Our model has a total of five parameters, all having an immediate and clear biological interpretation. 
We do not tune any of these parameters. Instead, we analyse experimental data to estimate parameter values that apply to skeletal muscle fibres of rhesus monkeys and use point estimates for our simulations (Figs. \ref{second_fig}.D, E).
In Fig.~\ref{second_fig}.D we show that stochastic simulations (Gillespie algorithm) model based on a net replicative advantage $k_{RA}$ for mutants predicts a travelling wave of heteroplasmy with a wave speed that is approximately 300 times faster than the observed speed.
Crucially, in Fig.~\ref{second_fig}E we show that stochastic simulations 
of our model for the same system predict a wave
of heteroplasmy with a speed of expansion comparable with observations. Setting up this simulation required some care, due to the slow dynamics caused by the large values of $N_{ss}$.  The wavefront of a travelling wave has the shape of a sigmoid whose steepness depends on its speed (see SI.\ref{steepness}), therefore one needs to initialise the wavefront with the correct steepness such that the wavefront remains approximately stationary. For all the other simulations presented in the paper, this is not an issue since the heteroplasmy wavefront assumes the steepness corresponding to its constant speed in a short time transient (faster dynamics), that can then be neglected. Instead, for Fig.~\ref{second_fig}E we established an appropriate approximate initial steepness after a series of trials, by verifying that the shape did not change appreciably after running the simulation for a sufficient number of days (around 50). The chosen initial heteroplasmy profile is  $h_0(x)=1/(e^{\tau x-b}+1)$ for $x>0$, with $\tau=7.15 \cdot 10^{-4} \mu m $ and $b=3.915$, over 500 regions of length $L= 30 \mu m$.  We also introduced approximations at the boundaries of the system. In order to simulate the effect of a macroscopic zone of the muscle fibre taken over by mutants we fixed the value of the heteroplasmy of leftmost region of the chain at 1.  At the right boundary, we truncated the values of heteroplasmy lower than $5 \cdot 10^{-3}$ to zero, to rule out the possibility that edge effects alter the speed of expansion of the heteroplasmy wavefront. We verified that the rightmost regions are mutant-free up to the longest simulated time. 
In  Fig.~\ref{second_fig}F, we show a wave of heteroplasmy that advances despite the mutants having a degradation rate $1\%$ higher than that of wildtype. This simulation has been performed with $N_ss={4}$ to obtain a faster dynamics and wave and avoid the issues with initialisation described above.

In addition, in Fig.~\ref{fourth_fig}A we provide evidence that our conclusions are robust to the uncertainty in parameter values. We have drawn parameter values according to the distributions inferred from experimental data (see discussion below for details). We have then obtained distributions of the wave speeds predicted by a SSD and RA model (FK waves), inserting parameter values in the corresponding formulae for wave speed, namely Eq. \eqref{eq:empirical_wavespeed} with the appropriate interpretation of $k$ for each case (setting $\beta=0$ for FK waves). The approach used to obtain the distributions of the predicted wave speeds in  is detailed in Ref. \cite{Johnston14Caladis} and can be easily reproduced using the online tool at \textcolor{blue}{http://caladis.org/} and the parameter estimates given below. The distributions plotted in Fig.~\ref{fourth_fig}A confirm that SSD is strongly favoured to account for the  expansion.

The parameters $\mu, \gamma, c, N_{ss}$ are common to SSD and 
to the RA model, whereas the net mutant replicative advantage $k_{RA}$ is replaced by $\delta$ in our model. We estimate the degradation rate $\mu$ from the half-life of mitochondria. Reported values for the half-life in the literature range from a few to 100 days \cite{Gross69,Neubert69,Diaz08, Burgstaller14, Johnston16}. We therefore assume that 
the half-life is uniformly distributed on the range (2,100) days. 
As a point estimate, we have chosen  10 days for the half-life, corresponding to a degradation rate of $\mu=0.07/$day.

$N_{ss}$ represents the average copy number per nucleus when only wildtypes are present. A study \cite{Miller03}
reports $N_{ss} \simeq 3500$ for healthy human skeletal muscle fibres. We therefore suppose that for rhesus monkeys muscle fibres our uncertainty on $N_{ss}$ has a
uniform distribution on the range (2000,5000). We have chosen 
$N_{ss} =3000$ since macaques are smaller animals. According to our empirical
formula in Eq. (\ref{wavesp_formula}), changing $N_{ss}$ from 3000 to 4000 would reduce the wave speed by about $10\%$, 
Therefore, we do not expect the uncertainty on $N_{ss}$ to affect our results significantly.

An important quantity in the study of muscle ageing, although not directly a
parameter of our model, is the internuclear distance $L$. 
It has been reported that in skeletal muscle of healthy mice around 30 nuclei are present in $1mm$ of fibre length, i.e. $\sim 1$ nucleus per $30 \mu m$ \cite{Bruusgaard03}. We therefore take $L=30 \mu m$ as a point estimate, while 
assuming a uniform distribution on the range $(25,35)\mu m$.

The hopping rate $\gamma$ is linked to the diffusion coefficient of mtDNA molecules in the muscle fibre, through the relationship $D=\gamma L^2$ \cite{Erban07}. 
Individual organelles in muscle fibres appear to be tightly 
packed. However, a study, Ref. \cite{Iborra04}, reported that the movement of nucleoids within organelles in these tissues can be effectively described as diffusion with a value of $D$
corresponding to $\gamma=0.1/$day. Therefore we choose this 
value as a point estimate of the individual mtDNA molecules. We quantify the  uncertainty about the value of $\gamma$ according to a uniform distribution on the range (0.05,0.15)/day 

The control strength $c$ is the only parameter for which we do not have an optimal measurement.
Our estimate is based on the fact that, in replicative cells, during a cell-cycle ($\approx$ 1 day)
the number of mtDNA roughly doubles. In this situation, in which a fast mitochondrial 
biogenesis is needed, we suppose that 
population is growing exponentially at a rate $\lambda(w=0, m=0)$. 
For a doubling time of 1 day, this corresponds to 
$$c=\frac{\ln2 /\textrm{day}-\mu}{N_{ss}}, $$
that for $N_{ss}=3000$ and $\mu=0.07/$day gives $ c=2 \cdot 10^{-4}/$day.
Since $c$ is not an independent parameter of our model, it is not present in the 
phenomenological wave speed formula Eq.~\eqref{wavesp_formula}.

The parameters $\delta$ and $k_{RA}$ correspond to the two ways of 
reproducing the increase in mutant carrying capacity (or density), respectively for our model and for a model based on a replicative advantage. From the experimental data \cite{Bua06} reported in Fig.~\ref{second_fig}B, in high-heteroplasmy regions density is approximately increased by a factor $\mathcal{F}=5$, as reported also in  Ref. \cite{Barthelemy01}. We therefore 
choose 5 as the point estimate of $\mathcal{F}$ and we assume that 
$\mathcal{F}$ has a uniform distribution on the range (2,8).
As for $\delta$, we notice that in our model the carrying capacity $m^*$ for mutants is obtained by  setting in Eq.~\eqref{linear_f_c_SI}   $\lambda = \mu$ and $w=0$ gives, leading to  $m^*=N_{ss}/\delta$. 
The carrying capacity for wildtypes
is obtained by setting $\lambda = \mu$ and
$m=0$ in  Eq.~(\ref{linear_f_c_SI}), leading to $w^*=N_{ss}$. Therefore,  
$\delta=1/\mathcal{F}$ and we chose  $\delta=1/5$ as a point estimate.
In a replicative advantage model, the mutants have a replication rate given by Eq.~\eqref{linear_f_c_SI} (with $\delta=1$) and the addition of a net replicative advantage $k_{RA}$, namely 
$$\lambda_m(w,m)=\mu + c(N_{ss}-w-m) +k_{RA}. $$ 
In order to reproduce an $\mathcal{F}$-fold increase of mutant carrying capacity, it must be $k_{RA}=cN_{ss}(\mathcal{F}-1$).
Indeed, setting $\lambda_m = \mu$ and $w=0$ yields $m^*=k_{RA}/c + N_{ss}$.
Hence, we assume $k_{RA}$ to be uniformly distributed on the range $(1,7)cN_{ss}$ and we choose $k=4cN_{ss}$ as a point estimate, yielding 
\begin{equation}
\lambda_m(w,m)=\mu + c(5N_{ss}-w-m)
\label{eq:mutant_repl_rate_RA}
\end{equation}
as the mutant replication rate in an RA model.

In Fig.~\ref{second_fig}F we show that stochastic survival of the densest yields a travelling wave of mutants even if mutants and wildtype replicate at the same rate and mutants are preferentially degraded. The mutant degradation rate is $10\%$ higher than wildtype, which is $\mu=0.07/$day. In this simulation, for numerical convenience, we have used $N_{ss}=4$, in order to obtain a faster dynamics. Other parameters are $\delta=2/3, \gamma=0.12/$day, $ c=0.4/$day. 

\section{Stochastic survival of the densest can account for mutational load in short-lived mammals} \label{ssod_short_lived}
In the penultimate Result section we have claimed stochastic survival of the densest lowers the value of  $R_{mut}$ required to yield a given mutant load through a neutral model. Here we explain and expand on this claim.

Previous efforts to understand the ability of neutral genetic models to account for mammalian ageing have often focused on understanding the compatibility between mutant loads and the de novo mutation rate ($R_{mut}$) through mathematical modelling. The approach is to develop a model predicting observed amounts of mutant molecules as a function of mutation rate and then choose the $R_{mut}$ that matches experimentally observed (i.e. target) mutant loads. In Ref. \cite{Elson01} a neutral stochastic model of degradation and replication of mtDNA successfully predicted commonly accepted values of mutant loads in humans with mutation rates $R_{mut}$ which fall into accepted ranges. However, it was later argued that this mechanism cannot explain mutant loads for short-lived animals like rodents \cite{KK13}, which show similar accumulation patterns on much shorter timescales ($\approx $ 3 years) \cite{Cao01,Herbst07}. For the model to predict the observed mutant loads in  short-lived animals, $R_{mut}$  must be so high that it would produce an unrealistically high mutational diversity, whereas experimental studies report a single deletion that clonally expands \cite{KK18}. It has therefore often been assumed that neutral stochastic models of mtDNA dynamics cannot explain the mutational load in short-lived animals.
 
However, previous neutral models neglected the spatial structure of muscle fibres and demanded the target mutant load be reached through pure random drift and continual addition of new sequence mutations. In this work we have shown that spatial structure induces a wave-like clonal expansion of mutants driven by stochastic survival of the densest. In Fig.~\ref{fourth_fig}B we have shown that this dramatically increases the chance that a mutation clonally takes over a macroscopic zone of a muscle fibre. The system simulated is a chain of $N=1001$ regions, each with $N_{ss}=4$, initialised with a single mutation mtDNA molecule with $\delta =0.1$ in the central region serving as a founder mutation and no additional mutational events. This creates a high-heteroplasmy front, of mean height $\simeq 0.30$, meaning that the mutants reach 100\% heteroplasmy  around $30\%$ of the time in an ensemble of $5 \cdot 10^2$ simulations. Early neutral models modelled muscle fibres as isolated, well-mixed populations of mitochondria \cite{Chinn99, Elson01}, which implies constant mean heteroplasmy  (red line in  Fig.~\ref{fifth_fig}B). For our example, with $1001 \times N_{ss} =4004$ mtDNA molecules, the initial value of heteroplasmy after one mutation arises, would be \nicefrac{1}{1004}, meaning that mutants would take over the  fibre around $0.025\%$ of the time. This simulation shows that spatial structure can dramatically increase the fixation probability of a founder mutation. The increase scales linearly with $N$, as it can be inferred by the previous example. Since a mice muscle fibre can contain up to thousands of nuclei (see next section), the fixation probability can be up to three orders of magnitude higher than previously thought.
Therefore, fewer mutations will need to arise (lower $R_{mut}$) before a high-heteroplasmy region manifests itself. In the next section we make this argument quantitative.

More generally, what determines  mutant load over time, and hence the progression of  sarcopenia, is not only $R_{mut}$,  as assumed in early studies,  but rather the speed of the wave of mutants. Different animals can have different mtDNA turnover rates, copy numbers, type of mutation and diffusion rate that, according to Eq. \eqref{wavesp_formula}, affect the wave speed. Therefore different species would be expected to experience sarcopenia on different time scales. In particular, short-lived  animals are predicted to have higher wave speeds for a larger proportion of fibres, since they have more glycolytic, low-copy-number fibres \cite{Pellegrino03}.

\section{Revisited estimates of mutant loads and mutation rate $R_{mut}$}
\label{revisited_loads}
In the main text (penultimate Results section) and in the previous section we have shown how a given $R_{mut}$ can yield a mutant load that is much higher than the predictions of previous models (Fig.~\ref{fourth_fig}B) that ignore the spatial structure of the system. 
Here we show that misinterpretation of the available experimental 
data -- caused by neglecting spatial structure -- has led to wrong estimates of the mutant loads. 

Most existing modelling studies \cite{Elson01, KK13, KK14} for humans and rodents 
have tried to match predictions to the following 
end-of-life target mutant loads:
5-10\% fibres must have reached a heteroplasmy 
level of ≥ 60\%. We remark that in these studies muscle fibres are modelled as a single region, with an unstructured and well-mixed mitochondrial population. In the following we refer 
to the the former quantity as \textit{threshold cell-fraction} (TCF) and to the latter as  \textit{threshold heteroplasmy} (TH).  
The TH of 60\% comes from a study on cybrid HeLa cells, stating 
that a 5196 base pair  deletion needs to reach a cell-fraction of $> 60\%$ 
before the cell shows a reduction in COX activity \cite{Hayashi91}.
However, TH is cell-dependent \cite{Hämäläinen13} and this value
needs not be relevant for skeletal muscle fibres in humans and rodents.
The TCF of 5\% \cite{Elson01, KK13} is often justified on the basis of
three studies \cite{Brierley98, Muller90, Cottrell00}. In the first study, 5\%  is the upper bound of the
range (0.1-5)\% \cite{Brierley98}. The second study found a maximum of 0.37\% 
COX-negative fibres in human limb muscle over a large age range \cite{Muller90}.
The last study \cite{Cottrell00} did observe high percentages of COX-negative muscle fibres ($\sim$ 40\%)
but, it refers to a patient affected by a mitochondrial deletion disease, whereas
we are interested in muscle ageing in healthy subjects.  


We propose new estimates
after examining data on skeletal muscle fibres of aged healthy mammals
\cite{Lopez00, Bua06, Moraes96, Brierley98} and considering the spatial structure of muscle fibres.  
In these studies, 
hundreds of serial sections of length $\sim 10 \mu$m were cut along muscle 
fibres and stained for cytochrome \emph{c} oxidase (COX) and succinate dehydrogenase (SDH) activities. 
Succinate dehydrogenase refers to complex II of 
the respiratory chain, the only complex fully encoded by the nucleus.
Therefore, negative SDH activities indicate a nuclear defect, whereas normal or 
hyperactive activities \emph{and} negative COX activities point towards mtDNA defects. An example of these data, from Ref. \cite{Bua06}, is Fig.~\ref{second_fig}A, where additionally the heteroplasmy is measured for each section. Notice that, for simplicity, we have reported the spatial dimension as continuous, while in reality the measurements (the blue circles in the plot) refer to the section as a whole.

First, we suggest a TH of $90\%$, based on the fact that
in Refs. \cite{Brierley98, Moraes96} regions of skeletal muscle fibres 
containing mitochondrial dysfunctions 
were shown to contain $> 90\%$ of deletion mutations.
Second, we claim that, in the light of a spatially-structured model,
the TCF is not the relevant quantity to use as a target for simulations. Instead, 
we propose a new quantity, the probability $P_{obs}(t)$ of observing an abnormal region in a fibre at age $t$, 
which we estimate using 
data in  Refs. \cite{Bua06, Lopez00}, based on the following argument.
For notational simplicity, we drop the time dependence, i.e. $P_{obs}=P_{obs}(t)$.
First, we make the following
assumptions: 
\begin{enumerate}

\item A fibre can be divided into regions, the building block of our model. A mutation can arise in a single region and can expand into neighbouring ones.

\item A mutation event cannot give rise to more than one abnormal zone. 
Although it may be that distinct but nearby abnormal zones in a single fibre are the result of a single mutation event, this was not observed in the studies we investigated.

\item An abnormal zone is caused by a single mutation event, not by more than one mutation that
expand and merge into a single zone.
\end{enumerate}
In Ref.~\cite{Bua06} segments of 2000 $\mu$m of 10652 
human vastus lateralis (VL) muscle fibres of aged (92 years old) subjects were analysed, 
and 98 abnormal zones  were found. In the context of our model and under our assumptions, 
this means that 98 mutations occurred \emph{and} expanded to $\sim$ 90\% 
in its original region as well as neighbouring ones, giving rise to a macroscopic abnormal zone. 
Based on our estimate of $L=30 \mu m$ for the inter-region spacing (SI.\ref{parameters}), 
the number of regions included in the 
measurements described above is $N = \frac{10652 \cdot 2000}{30}$, meaning that the probability 
that a mutation arises in any region and subsequently spreads in an observable manner is
$ 98/N \approx 1.4 \cdot 10^ {-4}$. This probability is equivalent to $P_{obs}$, the probability per region of
observing an abnormal zone.
In Refs. \cite{Lopez00}  segments of 1600$\mu$m of 2115 muscle fibres of 34-year-old
rhesus monkeys vastus lateralis (VL) were analysed, 
and 51 abnormal zones  were found. By the same reasoning, this corresponds to  $P_{obs}=0.05\%$.
These estimates for rhesus monkeys and humans refer to subjects at the end of the typical lifespan for 
the respective species, 
hence our proposed end-of-life $P_{obs}$ are in the range $\approx (1-5)\cdot 10^{-4}$.

Importantly, it is the spatial structure of our model, with its numerous regions composing a fibre, that
leads to the introduction of $P_{obs}$, whose estimates are orders of magnitude
lower than those for the TCF. We can see this with a hypothetical example.
Let us suppose that a study on human VL reports that $10\%$ of examined fibres of an individual of age $t$
show an abnormal zone  at some point along their length. For simplicity let us suppose that the deletion has fixed in this abnormal zone. Previous studies modelled fibres as a single well-mixed population. For a population of size $N_{ss}$, if we consider that each molecule has the same chance of fixing  given by $\nicefrac{1}{N_{ss}}$ as in the generalised Lotka-Volterra model, this requires that $R_{mut}t \approx 0.1$ (neglecting times to coalesce). This is because we expect $N_{ss} R_{mut}t$ mutations to arise in a time $t$. Let us now take into account the spatial extension of the fibres.
An average VL fibre of length $\approx 6.5$cm is constituted by $N=\frac{6.5 \times 10^4}{30} \approx 2200$ regions, assuming that a region has length $30 \mu m$, see SI.\ref{parameters}).
A correct interpretation of the experimental observation is that in at least one of these regions a mutation arises and fixes with probability 0.1. 
This corresponds to a value of $P_{obs}$ such that $1 -(1-P_{obs})^{2200} = 0.1$,
namely (for large $N$) $P_{obs}= \nicefrac{0.1}{N} \approx 4.5 \cdot 10^{-5}$. In Fig.~\ref{sixth_fig}B we show that in the context of stochastic survival of the densest a single mutation with 
$\delta=0.05$ has a chance of fixing $> \nicefrac{1}{N_{ss}} =0.25$. We have performed other simulations for $3 \leq N_{ss} \leq 20$ in the range and $ 0.05 <\delta < 0.5$, finding that in all cases the chance of fixing of a single founder mutation is $\approx \nicefrac{1}{N_{ss}}$. Therefore, we can conclude that in the spatially extended case $ R_{mut} t \approx  P_{obs} \approx  \nicefrac{0.1}{N} $, where we have assumed that $R_{mut}$ is so low that we can treat $R_{mut}t$  as a probability. We see that in this case the inferred mutation rate  is three order of magnitudes lower than in the well-mixed population context. More generally, 
for large $N$, which is true for muscle fibres, we predict $R_{mut}t$ to be $N$ times  smaller than previously assumed.

In summary,  we have presented two reasons why $R_{mut}$ could be a small (and so realistic) number. In this section we have argued that the literature has focused on the TCF but $R_{mut}$ does not need to be equal to this number -- it is, in fact, several orders of magnitude smaller. In the previous section we have shown that, because of stochastic survival of the densest, a single founder mutation spreads with a macroscopic probability, and this leads to values of $R_{mut}$ much lower than in the well-mixed population picture for a given mutant load.
In conclusion, neutral models based on random drift do not need an excessively high mutation rate to reproduce the observed mutant loads and can be a strong candidate to explain the expansion of mitochondrial deletions in muscle fibres.





\begin{figure}
\centering
\includegraphics[width=0.9\textwidth]{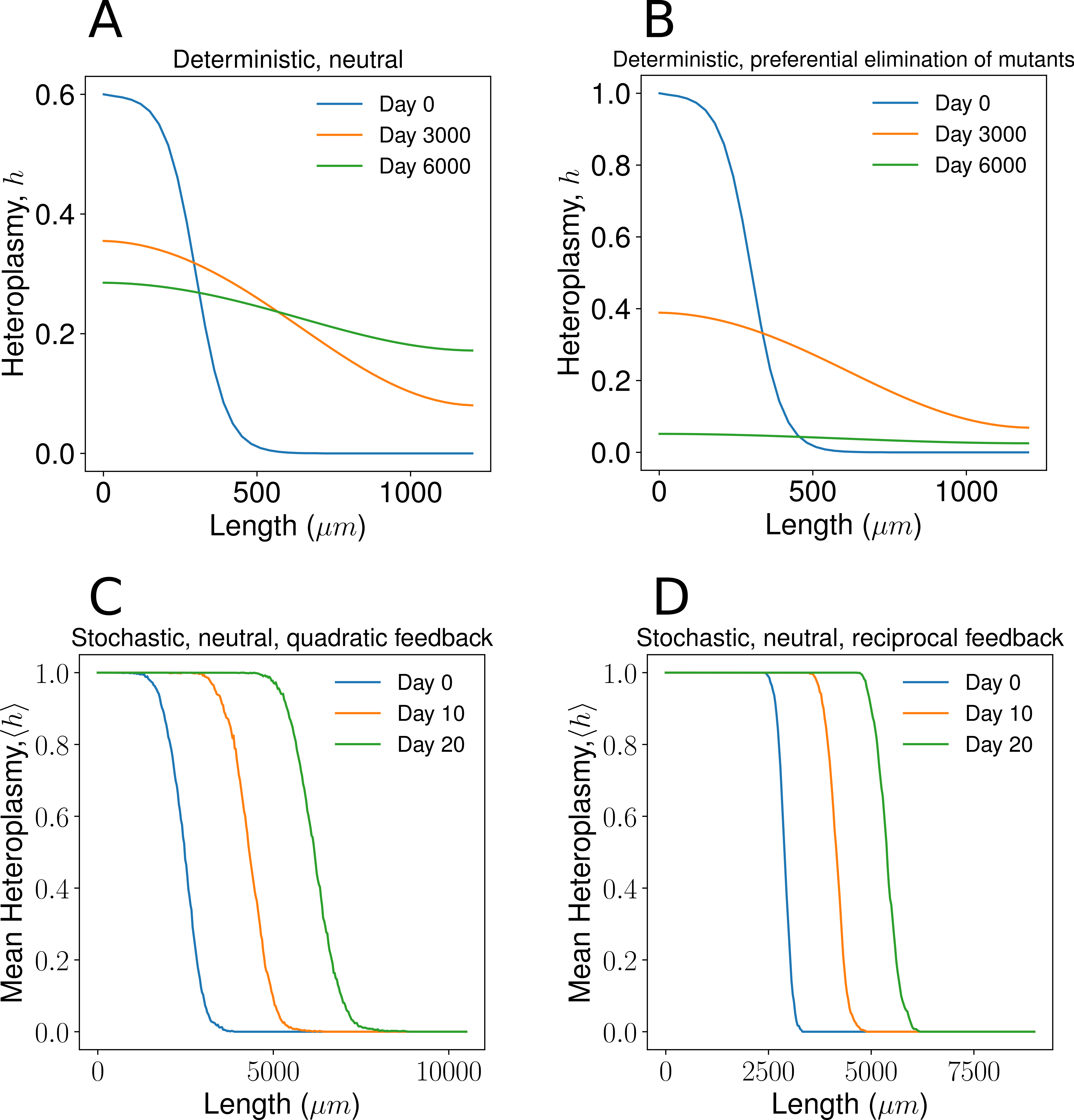}
\caption{\textit{A deterministic model deployed on a chain does not produce mutant expansion despite differences in carrying capacity, while different choices of feedback control still produce a wave of mutants when stochasticity is present}.
(A) The deterministic ODE model of Eq.~\eqref{ODE_system_2} applied to a chain of regions does not produce a travelling
wave of heteroplasmy, even with larger carrying capacity for mutants ($\delta<1$). Rather, we observe diffusion of the heteroplasmy front (with reflecting boundaries).
The curves are the heteroplasmy values obtained by ODE integration of a system 
analogous to Eq. (\ref{ODE_system_2}) for 41 regions. Parameters are $\mu=0.07$/day $\gamma=0.1$/day,  $\delta=0.1$, $c=0.02$/day, $N_{ss}=3000.$ (B) In the deterministic chain, when the degradation rate of mutants is higher than that of wildtype, mutants go extinct despite a higher carrying capacity. Parameters as in (A), with mutant death rate being $1\%$ higher than mutant. (C) Quadratic and (D) reciprocal feedback control produce a qualitatively similar wave-like expansion of mutants in a stochastic model. See 
SI.\ref{other_feedbacks} for definition of the feedbacks and parameter values. }
\label{seventh_fig}
\end{figure}

\section{Quadratic and reciprocal feedback produce a wave of heteroplasmy }
\label{other_feedbacks}
In Fig.~\ref{seventh_fig}C,D we show that other types of feedback controls  produce an analogous
travelling wave of heteroplasmy without the need of a replicative advantage for mutants.
The two additional types of controls are the \textit{quadratic}  control, namely
\begin{equation}
\lambda(w,m)=\mu + a[N_{ss}^2 - (w+\delta m)^2],
\label{eq:quad_feedback}
\end{equation}
and the \textit{reciprocal} control, given by
\begin{equation}
\lambda(w,m)=\mu - b\left (\frac{1}{N_{ss}} - \frac{1}{w+\delta m} \right ),
\label{eq:rec_feedback}
\end{equation}
where $a$ and $b$ are control strength parameters.
Importantly, these control mechanisms all produce 
wave speeds that decrease when $N_{ss}$ increases, a fundamental difference
between our model and models based on a net replicative advantage for mutants. 

The data reported in Fig.~\ref{seventh_fig}C,D are obtained averaging over 400 realizations with the following parameters.
For quadratic feedback, Fig.~\ref{seventh_fig}C and Eq.~\eqref{eq:quad_feedback}: $\mu=7 \cdot 10^{-2}$/day, $\gamma=0.12$/day,  $\delta=\nicefrac{2}{3}$, $a=0.4$/day, $N_{ss}=4.$   
For reciprocal feedback, Fig.~\ref{seventh_fig}D and Eq.~\eqref{eq:rec_feedback}: $\mu=7 \cdot 10^{-2}$/day, $\gamma=10^{-2}$/day, $\delta=0.1$, $b=1.5$/day,  $N_{ss}=3.$
The small values of  $N_{ss}$ are chosen for numerical convenience, since a smaller copy number produces a faster dynamics and faster waves.


\section{Relationship between steepness and speed of a travelling wave}
\label{steepness}
This section is based on Secs. 13.2 and 13.3 of Ref. \cite{Murray02}.
Let us consider the classical Fisher-Kolmogorov reaction-diffusion equation
\begin{equation}
\frac{\partial h}{\partial t}=kh(1-h) + D \frac{\partial^2 h}{\partial x^2},
\label{FK_eq}
\end{equation}
where $h$ is heteroplasmy, $k$ is the reaction rate and $D$ is the diffusion coefficient. 
This equation admits travelling wave solutions moving in the positive $x$-direction  
with speed $c$ given by
$$h(x,t)=h(z)=h(x-ct),$$
having defined $z=x-ct$. 
If the mutants of our system had a replicative advantage $k$, Eq. (\ref{FK_eq}) would describe the 
wave-like expansion of the high-heteroplasmy front in the continuous-space approximation.
If the system is initialised such that 
$$h(x,0) \sim e^{-ax}  \hspace{0.5cm} \text{as}  \hspace{0.5cm} x \rightarrow \infty, $$
for arbitrary $a>0$,
the heteroplasmy wavefront will evolve into an approximately sigmoidal shape with a steepness 
$\tau$ linked to the wave speed through
\begin{equation}
c=\frac{k}{\tau},
\end{equation}
meaning that the slower the wave, the steeper the wavefront (larger $\tau$).

\begin{figure}
\centering
\includegraphics[width=0.5\textwidth]{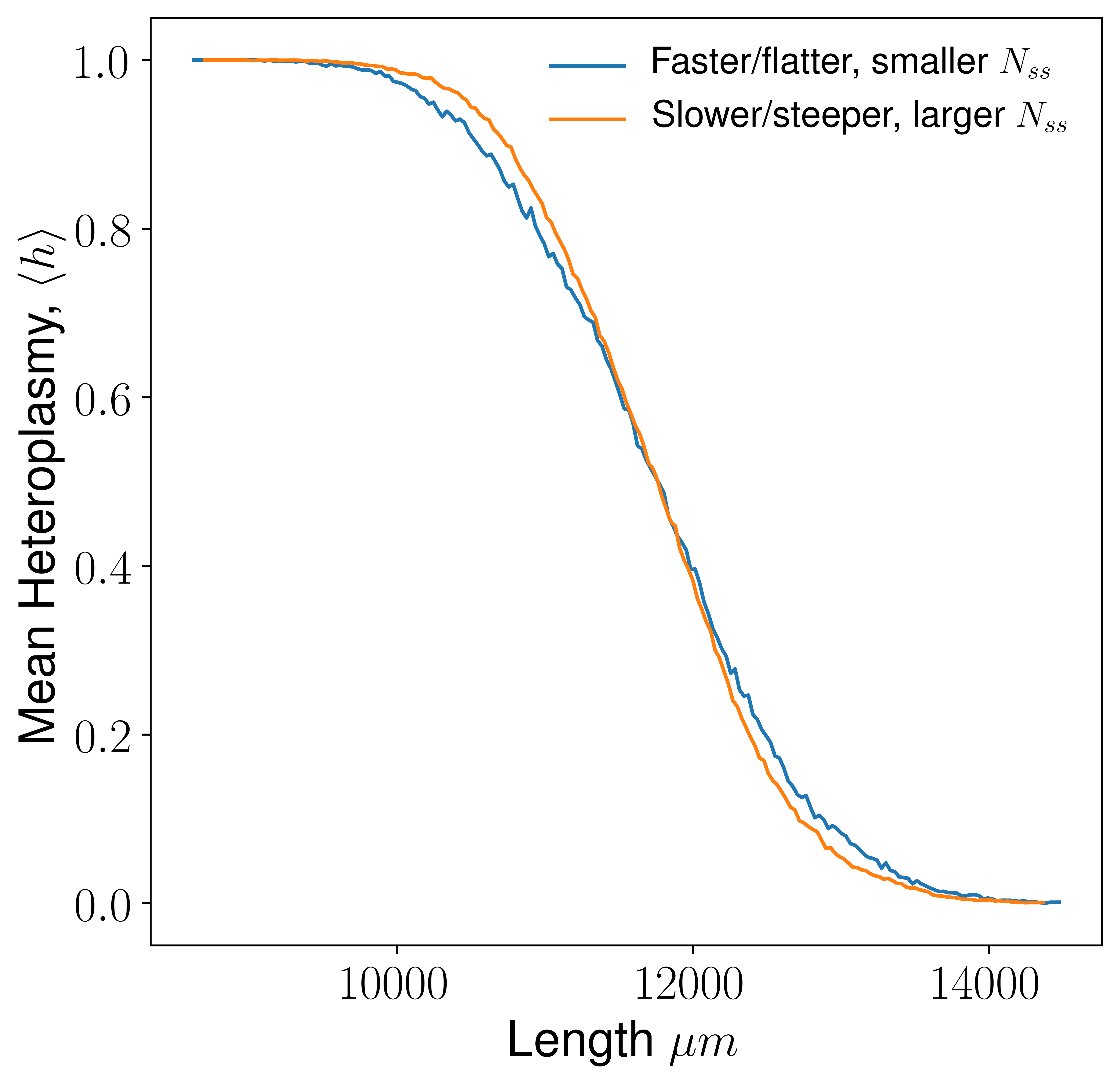}
\caption{The two curves are the travelling wavefronts of two systems that differ only in the average wildtype copy number
 $N_{ss}$. The smaller the $N_{ss}$, the faster the wave, according to Eqs.~\eqref{eq:empirical_wavespeed} and \eqref{eq:effective_reac_rate}. The blue curve is the wavefront propagating in the system with the  smaller
$N_{ss}^{fast}=2$, while the blue curve is relative to $N_{ss}^{slow}=6$.  The plot shows that faster waves have a flatter wavefront,
as it is the case for FK waves.  }
\label{fig:steepnesses}
\end{figure}

We speculate that waves driven by stochastic survival of the densest can be described by an equation analogous to 
Eq. (\ref{FK_eq}), with an effective reaction rate induced by stochastic survival of the densest, that we hypothesise is given by Eq. (\ref{eq:effective_reac_rate}).
Therefore, also for our model we expect that steeper waves will be slower. In absence of a formal proof, we have
verified this intuition computationally. In Fig.~\ref{fig:steepnesses} we plot the travelling waves of heteroplasmy 
for two systems that differ only in copy number $N_{ss}$. The blue curve corresponds to a system with a smaller
$N_{ss}$ and hence, according to Eq. (\ref{wavesp_formula}) presenting a faster wave; 
the orange curve corresponds to a larger  $N_{ss}$ and hence a slower 
wave. As expected, the slower wave has a steeper wavefront.

In Fig.~\ref{third_fig} we have reported experimental data on human (panels A and B) and rat (panels C and D) muscle fibres, 
showing that the steeper wavefront corresponds to the  fibre with a larger $N_{ss}$. 
We can conclude that the steeper wavefront corresponds to a
slower wave. Hence, the data reported in  Fig.~\ref{third_fig} support the prediction 
of our model that slower waves travel in fibres with a larger $N_{ss}$.

\section{Rhesus monkeys data collection}
\label{sec:monkeys_speed}
The data plotted in Fig.~\ref{second_fig}C on the length of the abnormal zones in muscle fibres of rhesus monkeys was originally published in Ref. \cite{Lopez00}. The measurements were performed on skeletal muscle tissues of 11 different animals and the data set consists of the lengths of several abnormal zones for each subject. For each animal, we only use the length of the longest abnormal zone, that we assume has originated from a mutation at birth. We consider shorter regions to be the results of mutations that have arisen later in life. Although these assumption are unlikely to be precisely true, our analysis is enough to give an indicative order-of-magnitude estimate, that can be considered as a moderately tight lower bound on the speed of expansion.

\section{Statistical methods}
Error bars in Figs.~\ref{first_fig}D and \ref{fifth_fig}A, B are standard error of the mean (SEM). The parameters of the fits reported in Figs.~\ref{second_fig}C and \ref{sixth_fig}C are estimated via linear least squares and the uncertainty reported is the standard deviation. For the fits reported in Figs.~\ref{second_fig}A and \ref{third_fig}A, C non-linear least squares is used and the uncertainty  is the standard deviation. The statistical test performed on the data in \ref{third_fig}B, D is a one-tailed unequal-variances \textit{t}-test (Welch's test). The wave speeds relative to Figs.~\ref{second_fig}D, E and those corresponding to the points in Fig.~\ref{sixth_fig}A have been calculated as explained in SI.\ref{pheno_wave_speed}.

\newpage

\bibliographystyle{ieeetr}
\bibliography{biblio}

\end{document}